\documentclass[12pt, clsnofoot, onecolumn]{IEEEtran}
 \pdfoutput=1
\usepackage[pdftex]{graphicx}
\usepackage{enumerate}
\usepackage{caption}
\usepackage{subcaption}
\usepackage{amsthm}
\usepackage{amssymb}
\usepackage{dsfont,physics}

\usepackage{setspace}
\usepackage[linesnumbered,ruled,vlined]{algorithm2e}
\usepackage{comment}
\usepackage{color}
\usepackage{float}
\usepackage{gensymb}
\usepackage{comment}
\usepackage{xcolor}
\usepackage{cuted}
\usepackage{soul}

\usepackage{cite}

\theoremstyle{plain}

\newtheorem{lemma}{Lemma}

\newtheorem{assumption}{Assumption}

\theoremstyle{definition}

\theoremstyle{remark}
\newtheorem{remark}{Remark}

\usepackage{mathtools} 

\newcommand{\expect}{\operatorname{\mathbb{E}}\expectarg}
\DeclarePairedDelimiterX{\expectarg}[1]{[}{]}{%
  \ifnum\currentgrouptype=16 \else\begingroup\fi
  \activatebar#1
  \ifnum\currentgrouptype=16 \else\endgroup\fi
}
\newcommand{\prob}{\operatorname{\mathbb{P}}\probarg}
\DeclarePairedDelimiterX{\probarg}[1]{(}{)}{%
  \ifnum\currentgrouptype=16 \else\begingroup\fi
  \activatebar#1
  \ifnum\currentgrouptype=16 \else\endgroup\fi
}
\newcommand{\innermid}{\nonscript\;\delimsize\vert\nonscript\;}
\newcommand{\activatebar}{%
  \begingroup\lccode`\~=`\|
  \lowercase{\endgroup\let~}\innermid 
  \mathcode`|=\string"8000
}

\def\*#1{\mathbf{#1}}
\def\&#1{\mathcal{#1}}
\def\.#1{\boldsymbol#1}
\def\^#1{\hat{#1}}
\def\[#1{\left #1}
\def\]#1{\right #1}

\DeclareMathOperator*{\argmin}{arg\,min\ }
\DeclareMathOperator*{\argmax}{arg\,max\ }

\DeclareMathAlphabet\mathbfcal{OMS}{cmsy}{b}{n}

\begin{document}
%\pagenumbering{gobble}  %this gets rid of page numbers
\bstctlcite{BSTcontrol}
\doublespacing

\title{Active and Dynamic Beam Tracking Under Stochastic Mobility\\
}

\author{\IEEEauthorblockN{Nancy Ronquillo,  and Tara Javidi}

\IEEEauthorblockA{Department of Electrical and Computer Engineering \\
University of California, San Diego \\
Email: \{nronquil,tjavidi\}@ucsd.edu}
}

\maketitle

\begin{abstract}% Communications at millimeter wave (mmWave) frequencies or above aim to meet the demand for higher data rates by using highly directional beams with access to larger bandwidth. An inherent challenge is tracking channel state information (CSI) necessary for mmWave transmission under systems with unpredictable mobility. 
 We consider the problem of active and sequential beam tracking at mmWave frequencies and above. We focus on the dynamic scenario of a UAV to UAV communications where we formulate the problem to be equivalent to tracking an optimal beamforming vector along the line-of-sight path. In this setting, the resulting beam ideally points in the direction of the angle of arrival with sufficiently high resolution. Existing solutions account for predictable movements or small random movements using filtering strategies or by accounting for predictable mobility but must resort to re-estimation protocols when tracking fails due to unpredictable movements. We propose an algorithm for active learning of the AoA through evolving a Bayesian posterior probability belief which is utilized for a sequential selection of beamforming vectors. We propose an adaptive pilot allocation strategy based on a trade-off of mutual information versus spectral efficiency. Numerically, we analyze the performance of our proposed algorithm and demonstrate significant improvements over existing strategies.
\end{abstract}
%\begin{IEEEkeywords}
%component, formatting, style, styling, insert
%\end{IEEEkeywords}
%\color{blue}

\section{Introduction}

A promising approach to meet the increasing demand for higher data rates is communication at Millimeter Wave (mmWave) frequencies and above due to the larger available spectrum resources. However, to mitigate the higher pathloss at these high frequencies antenna arrays with many elements must be utilized to overcome propagation and atmospheric losses by concentrating power through directional beamforming \cite{Maccartney2013,Rappaport2017, Molisch2017}. This means that the feasibility of sub-mmWave communication directly depends on robust directional beamforming with fast and near real-time acquisition of channel state information (CSI).

For communication with static, or quasi-static channel conditions, the problem of acquiring CSI is only necessary initially for a given coherence time in order to establish beam alignment. Many innovative solutions have been proposed to obtain robust beamforming for communication, even at a low SNR regime ($<5$ dB)  \cite{Giordani2016,Heath_Overview,Abari2016, Song_TWC2018, Alkhateeb2014, ChiuJSAC2019, Ronquillo19}. 
Among existing solutions, those strategies with the first response time, or equivalently shortest initial access piloting phase, leverage a beamforming codebook with pseudo random beam sweeping such as \cite{Abari2016,Song_TWC2018} or a hierarchical beamforming codebook \cite{Alkhateeb2014,ChiuJSAC2019, Ronquillo19}. In particular, our prior work \cite{ChiuJSAC2019, Ronquillo19} has shown that sequential selection of the beamforming vectors reduces the expected number of measurements $\mathbb{E}[\tau_{max}]$ required in order to establish reliable communication, where the benefits over passive or random approaches \cite{Abari2016,Song_TWC2018, Alkhateeb2014} are greater in the low SNR regime ($<5$~dB).  

The problem becomes far more challenging under dynamic channel conditions such as a cellular enabled mobile unmanned aerial vehicle (UAV) systems, where a transmitters mobility impacts the path angle of arrival (AoA) at the receiver and vice-versa. 
CSI acquisition for maintaining beam alignment over time in the scenarios of high mobility, commonly referred to as beam tracking, has been extensively studied in \cite{Xiao2016,Va2016_GSIP,Zhao2017,GaoTVT2017,  shaham2019extended,Yang2019,Huang2020,Wang2020_Mobicom,Zhang_UAVsurvey, Mozaffari_UAVtutorial}. 
%ZhangTWC2019,
%We restrict our attention to the cases of beamforming under the practical implementation constraints of a beamforming codebook, where the challenge can be thought of as effective and dynamic joint optimization of beam and pilot selection. 
Existing proposals \cite{Va2016_GSIP,Zhao2017,GaoTVT2017, shaham2019extended,Yang2019,Huang2020,Wang2020_Mobicom} for handling very high mobility rely heavily on schemes of pilot allocation, switching between data transmission and pilot phases. More specifically, the pilot phase is generally used for CSI and/or mobility estimation, for example by leveraging compressive sensing or least squares techniques \cite{Zhang_UAVsurvey}, while data transmission depends on highly reliable estimates of the best predicted beam.

In contrast, we propose a method for active and dynamic learning of the AoA throughout both the data transmission phase as well as pilots. Specifically, our method actively selects beamforming vectors based on evolving a Bayesian probability belief. In the absence of excess uncertainty and with moderate SNR, posterior updates rely on the signal energy (which allows the update to be agnostic to the knowledge about the data sequence). When the belief displays a large variance, in contrast, our algorithm deploys a pre-designed pilot sequence to speed up learning. The adaptive pilot selection relies on analysis of the trade-off between the mutual information versus the spectral efficiency.

\subsection{contributions}
% %In this work, consider the problem of beamforming and tracking at the RX for a channel dominated by the LoS path. 
% We formulate the problem to be equivalent to the problem of acquiring and maintaining robust point estimates of the AoA at the RX, i.e. tracking of the AoA.
We consider the practical implementation with a single RF chain, and small scale channel dominated by the line-of-sight (LoS) single path, where CSI acquisition reduces to the problem of estimation and tracking of the dynamic angle of arrival (AoA).

Many Proposed solutions in the literature focus on tracking predictable movements, such as a UAV moving at a known or estimated velocity, or where the AoA or UAV trajectory and position can otherwise be inferred from the geometry \cite{ GaoTVT2017, Huang2020}. These solutions rely on a channel estimation phase to lock into an initial estimate and track with beams based on these estimates thereafter. Complimenting this approach, Kalman filtering based strategies for estimating the AoA can support tracking in face of Gaussian error in predicted movement \cite{Va2016_GSIP, Zhao2017}- possibly supplemented with geometric calculations for predicting the position of the transmitter instead \cite{shaham2019extended}. For largely unpredictable movements, such as large jumps or changes in trajectory, these solutions operate model-agnostic and handle sudden changes by implementing a form of adaptive switching between estimation and tracking based on either recurringly allocating pilots \cite{Yang2019, Wang2020_Mobicom} or by constantly evaluating the quality of tracking \cite{Va2016_GSIP, Huang2020}.
The most recent works have studied the benefits of using beams covering wider angular regions, rather than exclusively using narrow beams, in order to capture both fast angle variations and reduce the pilot overhead \cite{Yang2019, Wang2020_Mobicom}. The pilot overhead can be reduced further by focusing locally according to current estimates, the caveat is reduced link quality due to the wider beam width.

In this work, we view the problem of beam tracking as an active and dynamic learning of the AoA under the caveat of measurement-dependent noise, where the signal and statistics are dictated by the type of measurement.
The problem of static search with measurement dependent noise has been studied from an information theoretic perspective where many works have established a connection to the problem of channel coding over a binary input channel \cite{Giordani2016,Chiu2016,Ronquillo2016,Lalitha2017,Kaspi2018}. Existing adaptive strategies for measurement selection based on posterior matching have been shown to provide strong information theoretical performance guarantees \cite{ChiuJSAC2019, Kaspi2018,Lalitha2017}. We draw on these works, leveraging the connection to channel coding, as well as our earlier work on real-time joint-source channel coding in \cite{Javidi2013_isit}, to develop our adaptive beamforming algorithm based on posterior matching using a beamforming codebook. Our work generalizes earlier work on beam tracking under predictable movement \cite{Huang2020}, as well as filter-based estimation strategies \cite{Va2016_GSIP, Zhao2017,shaham2019extended} for small angle variations. 
Our contributions are as follows:

\begin{enumerate}

\item\textbf{Active and Sequential beam selection} We propose a method for sequentially selecting beamforming vectors in an active manner from a pre-designed hierarchical beamforming codebook, with beams of various widths and variable achievable gains based on posterior matching \cite{Kaspi2018}. Our methodology consists of a) matching the selection of the beam to the posterior belief about the AoA and b)evolving the posterior via Bayesian updates (predictive filtering) in order to incorporate mobility information. %We provide closed form solutions of the posterior update and prediction equations for three Markov movement models that represent predictable and unpredictable mobility examples.

\item\textbf{Adaptive Pilot Allocation} We propose an adaptive pilot allocation strategy to complement the communication scheme. Specifically, we propose to trade-off the pilot enabled channel estimation and data transmission phases by characterizing and balancing the mutual information about the AoA against the achievable spectral efficiency. Combined with our active and sequential beam selection and dynamic tracking of the posterior, our approach trades off the pure exploration of pilot transmission against exploitation in the data transmission, as is often done in reinforcement learning \cite{RL_book}.
% The exploitation, hence enables 
% is a special case of the exploitation vs. exploration dilemma, which enables some learning about the AoA in the data transmission phase and is rooted in the reward optimization learning paradigm of reinforcement learning algorithms \cite{RL_book}.

\item\textbf{Numerical Results} We demonstrate via simulations the superior performance of our proposed communication scheme over prior work in terms of normalized beamforming gain and pilot overhead under three stochastic mobility models. In particular, we first consider a predictable movement scenario for which we recover our prior work on initial beam alignment \cite{ChiuJSAC2019}, and match the performance of prior work \cite{Huang2020,Va2016_GSIP}. We then consider the cases where 1) the AoA prediction has large mean square error, modeled by a zero-mean Gaussian noise term, or 2) is subject to Bernoulli angular jumps with a known bias.
Comparing our work against the algorithms of \cite{Huang2020,Va2016_GSIP,Yang2019}, we demonstrate our robust beamforming and efficient tracking of the AoA with minimum pilot overhead. 
In practical terms, and under stochastic mobility, this means that our algorithm achieves significantly higher average beamforming gains and reduced pilot overhead.

\end{enumerate}

\subsection{Notations} We use boldface letters to represent vectors or matrices. $\|\textbf{A}\|_0$ is the $l_0$ norm, i.e. sum of non-zero entries, while $\|\textbf{A}\|$ is the $l_2$ norm of $\textbf{A}$. For a number $c = a+ib$, $\|c\| = \sqrt{a^2+b^2}$ is the complex modulus. We denote the space of probability mass functions on set $\mathcal{X}$ as $P(x)$.
%We denote the Kullback-Leibler (KL) divergence between distribution $P$ and $Q$ by $D(P \| Q) = \sum_x P(x) \log \frac{P(x)}{Q(x)}$. The mutual information between random variable $X$ and $Y$ is defined as $I(X,Y) = \sum_{x,y} p(x,y) \log \frac{p(x,y)}{p(x)p(y)}$, where $p(x,y)$ is the joint distribution and $p(x)$ and $p(y)$ are the marginals of $X$ and $Y$. 
Bern$(q)$ is the Bernoulli distribution with parameter $q$. $\mathcal{N}(\mu,\sigma^2)$ is the Gaussian distribution with mean $\mu$, and variance $\sigma^2$.
%, and Bern$(x;p) = p^x (1-p)^{1-x}$. %Let $I(q;p)$ denote the mutual information of the input $X \sim$ Bern$(q)$ and the output $Y$ of a BSC channel with crossover probability $p$. $C_1(p) := D( \text{Bern}(p) \| \text{Bern}(1-p) )$ denotes the error exponent of hypothesis testing of $\text{Bern}(p)$ versus $\text{Bern}(1-p)$. 
$\mathcal{CN}(\tilde{\mu}, \sigma^2)$ denotes the complex circularly symmetric Gaussian distribution, where the probability density function for $\tilde{\mu}\in \mathbb{C}$ and $\tilde{x}\in \mathbb{C}$ is given as $g_{\tilde{\mu}}(\tilde{x})= \frac{1}{\pi \sigma^2} e^{-\frac{\|\tilde{x}-\tilde{\mu}\|^2}{\sigma^2}}$ for real and imaginary parts with variance $\frac{\sigma^2}{2}$.
$\chi^2(k, \lambda)$ denotes the non-central chi-squared distribution with $k$ degrees of freedom, non-centrality parameter $\lambda$, and scaled by $\sigma^2$ where the probability density function is given as $c_{\lambda}(x)=\frac{1}{\sigma^2}e^{-(\frac{x}{\sigma^2}+\frac{\lambda}{2})}\sum\limits_{k=0}^{\infty}\frac{(\frac{x\lambda}{2\sigma^2})^k}{(k!)^2}, \ x\geq 0$.
$\Re{c}, \Im{c}$ denote the real and imaginary parts of a complex number $c$, respectively. 

\begin{figure}
    \centering
    \includegraphics[width = 0.65\textwidth]{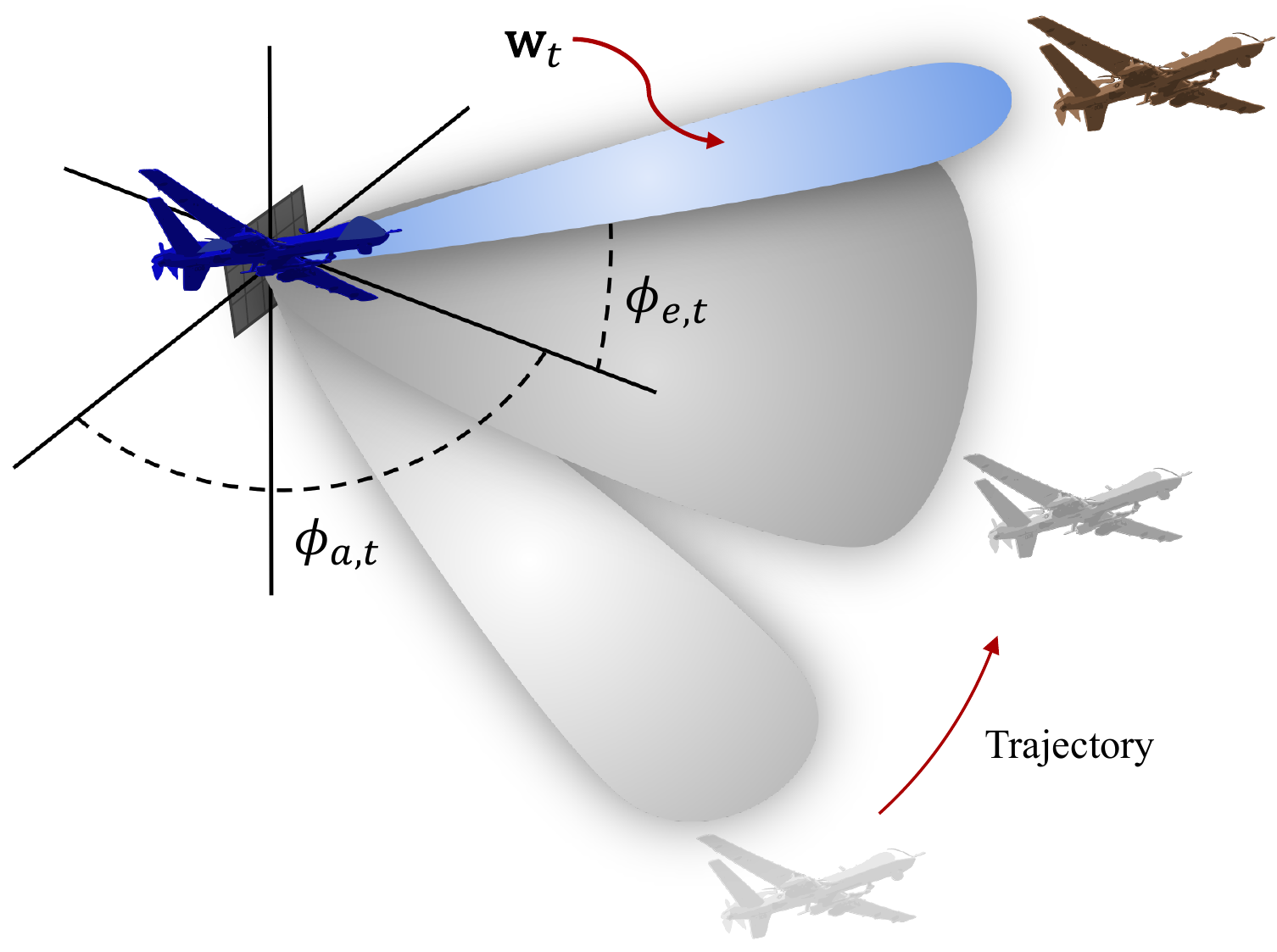}
    \caption{UAV beamforming setup for AoA $\Phi_t = (\phi_{a,t}, \phi_{e,t})$.}   
    \label{Moving_phi_mobjour}
\end{figure}
\section{System Model}
Consider a UAV to UAV communication set up with adaptive processing for a receiving UAV (RX) with a Uniform Planar Array equipped with $N\times M$ antennas and 3-D angular range.\footnote{Beamforming of the receiver can be done with reduced reliance on feedback, and hence is chosen here for its simplicity. We note that the proposed algorithm is also suitable for a 2-D set-up with a  Uniform Linear Array and $N$ antennas as discussed in Sect.~\ref{allexamples_mobjour}.} On this section we first provide the system model at a given time slot $t$. Fig.~\ref{Moving_phi_mobjour} illustrates the beamforming setup for an RX tracking a mobile TX UAV across its movement trajectory. A transmitting UAV (TX) has fixed beamforming acting as a single virtual antenna. We consider a low-power set-up where both UAVs use a single RF Chain. 

\subsection{Signal Representation}\label{signal_mobjour}
The RX combines the signal from the antenna elements to the RF chain by the directional beamforming vector $\mathbf{w_t}\in \mathbb{C}^{NM}$ at $t = 1,2, \ldots, T$, where $t$ represents a beamforming or sampling time slot. Without loss of generality, we assume normalized beamforming vectors such that, $\|\*w_t\|^2=1$.

For an air-to-air communication scenario a transmitting UAV will likely be unobstructed and free of reflectors, where we may assume that communication is dominated by the line-of-sight path. Therefore, we use the stochastic multi-path model (see Ch.7 in \cite{Tse2005}) with assumption of a single dominant path.
\begin{assumption}\label{assum:singlepath_mobjour}
The small-scale channel can be described by an $NM\times1$ complex vector:
\begin{equation} \label{eq:singlepath_mobjour}
    \*h = \alpha_t \*a(\Phi_t),
\end{equation}
where $\alpha_t \in \mathbb{C}$ is the complex path gain, 
\begin{equation}
\begin{aligned}
    \*a(\Phi_t) :=  &\sqrt{\frac{1}{NM}}\Big[ 1, e^{j\frac{2\pi d }{\lambda} [\sin{\phi_{a,t}}\sin{\phi_{e,t}}+\cos{\phi_{e,t}}]}, e^{j\frac{2\pi d }{\lambda} [(N-1)\sin{\phi_{a,t}}\sin{\phi_{e,t}}+(M-1)\cos{\phi_{e,t}}]\Big]}
\end{aligned}
\end{equation}
\label{singlepath_alpha_mobjour}
is the array manifold at the receiver with antenna spacing $d$, and $\Phi_t = (\phi_{a,t}, \phi_{e,t})$ is the AoA in azimuth and elevation for $\phi_{a,t}\in [\theta_{\text{min}},\theta_{\text{max}}]$ and $ \phi_{e,t}\in [\psi_{\text{min}},\psi_{\text{max}}]$.
\end{assumption}
% When different UAV are simultaneously transmitting, the detected signal from a given UAV, $k\in K$, can be written as
% \begin{equation}
% \begin{aligned}
%     y_t^k &{=} \alpha_t \*w_t^H (\sum_{k'=1}^K \*h_{k'}  \mathbf{s}_{k'}^{T} ) \mathbf{s}_k^{*} +\*N_{t} \mathbf{s}_k^*.\\
% \end{aligned}
% \end{equation}

% We consider communications from a TX to the RX over one beamforming slot to be characterized by one of two possible phases: pilot training or data transmission. Let the current phase be denoted by  $e_t \in \{P, D\}$. The continuous time received signal over beamforming slot $t$ is 
% \begin{equation}
% \begin{aligned}
%     r_t(\tau) &{=} \alpha_t \*a(\Phi_t)  \mathbf{x}(\tau)\\
% \end{aligned}
% \end{equation}
% where $\*x(\tau)$ is the transmitted sequence $
%     \*x(\tau) = \sum_{n=1}^{Nc}\sqrt{P_T}x_n p_r(\tau-nTc)$, 
% consisting of transmitted symbols $x_n$ and a pulse shaping function $p_r(\tau)$ where $\int |p_r(\tau)|^2 d_{\tau}=1$, for $N_c$ modulation symbols with $T_c$ symbol duration. 
A receive beamforming vector $\*w_t$ is applied for the duration of the beamforming slot. The discrete time baseband representation of the received signal is
\begin{equation}\label{genobsv_mobjour}
\begin{aligned}
    y_t &=\sqrt{P_T} \*w_t^H \*h x_t + \*w_t^H\*n_{t}.\\
\end{aligned}
\end{equation}
where $x_t\in \mathbb{C}$ is the modulated complex symbol\footnote{In practice, a beamforming slot may be formed by a block of $Q_x\geq 1$ transmitted symbols. For ease of exposition, we assume a beamforming slot corresponds to a single symbol $Q_x=1$, however a larger block length is a straightforward extension.}, and $\mathbf{n}_{t}
\sim \mathcal{CN}(0_{[NM \times 1]}, \sigma^2\*I)$ is the additive AWGN. We consider perfect knowledge of the operating SNR, defined as $\frac{P_T}{\sigma^2}$ which is the SNR that would be received with narrowest aligned beamforming.    
% \begin{equation}
% \begin{aligned}
%     y_t &{=} \alpha_t \sqrt{P_r}\*w_t^H \*a(\Phi_t)  \mathbf{x}_t +\*w_t^H\*N_{t} .\\
% \end{aligned}
% \end{equation}

% For an applied receive beamforming vector $\*w_t$, the received signal in one beamforming slot is 
% \begin{equation}
% \begin{aligned}
%     y_t &{=} \alpha_t \sqrt{P_r}\*w_t^H \*a(\Phi_t)  \mathbf{x}_t +\*w_t^H\*N_{t} .\\
% \end{aligned}
% \end{equation}
% where $\mathbf{x}_t \in  \mathcal{C}^{Q\times 1}$ is the transmitted sequence and $\*N_{t}\sim \mathcal{CN}(0_{[NM \times Q]}, \sigma^2\*I)$ is the spatially uncorrelated AWGN. 

\begin{remark}\label{fixedalpha_mobjour}
We note that given knowledge of the operating SNR $\frac{P_T}{\sigma^2}$, we can assume $P_T=1$ without loss of generality. Further more, we assume known fading $\alpha_t=1$. While the extensions detailed in \cite{Ronquillo19} can handle stochastic and time varying complex gain through simultaneous estimation of $\alpha_t$, we leave this as an area of future work. 
\end{remark}
\subsubsection{\textbf{Receive a pilot $(e_t=P)$}}
During the pilot training phase, denoted by $e_t = P$, the transmitted symbols $x_t$ are assumed to be known at the receiver. Thus, the discrete time detected pilot signal for a beamforming slot $t$ can be expressed as 
\begin{equation}
\begin{aligned}\label{eq:obsv_mobjour}
    Z_t(P) &=y_t\frac{x_t^{\ast}}{\|x_t\|^2}\\
    &=\*w_t^H \*a(\Phi_t)  + \*w_t^H\*n_{t}.\\
\end{aligned}
\end{equation}
Note, for a pilot where $\|x_t\|^2=1$ and $\|\*w_t\|^2=1$ $\eta_t=\*w_t^H \mathbf{n}_{t}x_t^{\ast}\sim \mathcal{CN}(0, \sigma^2)$. This means that $Z_t(P)$ is circularly symmetric complex Gaussian random variable with variance $\sigma^2$, i.e. the observation $Z_t(P)\sim\mathcal{CN}(\*w_t^H \*a(\Phi_t), \sigma^2)$. The conditional probability is
\begin{equation}\label{condprob_pilot_mobjour}
f_{Z_t(P)|\Phi_t, \*w_{t}}\big( \xi_t\big| (\theta_i,\psi_j), \*w_{t}\big) = g\big(\xi_t;G_{i,j}\big),
\end{equation}
where $G_{i,j} = \*w_t^H \*a(\theta_i,\psi_j)$
is the gain conditioned on $\Phi_t=(\theta_i,\psi_j),$ and beamforming vector $\*w_{t}$.

\subsubsection{\textbf{Receive data} $(e_t=D)$}
On the other hand, in the data transmission phase the transmitted data is unknown. 
We apply an additional processing step to the discrete time signal (\ref{genobsv_mobjour}), where we calculate the received power (which will be used for the purpose of our tracking algorithm):
\begin{equation}
\begin{aligned}\label{eq:obsv_data_mobjour}
    Z_t(D) &=\|y_t\|^2\\
    &=\|\*w_t^H \*a(\Phi_t)x_t + \*w_t^H\*n_{t}\|^2.\\
\end{aligned}
\end{equation}
More precisely, the next lemma provides the distribution of $Z_t(D)$ conditioned on AoA $\Phi_t$ and beamforming vector $\*w_t$: 
\begin{lemma}\label{cond_prob_data_mobjour}
Let each transmitted symbol $x_t\in \mathbb{C}$ have minimum energy $\|x_t\|^2\geq1$. Then, conditioned on $\Phi_t=(\theta,\psi)$ and beam vector $\*w_{t}$, $Z_t(D)$ follows a scaled non-central chi-squared probability distribution function $Z_t(D)\sim\chi^2(k, \lambda_t)$ with $k=2$ degrees of freedom and time-varying non-centrality parameter $\lambda_t=\frac{2\|\*w_t^H \*a(\theta,\psi)\|^2}{\sigma^2}$
\begin{equation}
f_{Z_t(D)|\Phi_t, \*w_{t}}\big(\xi_t\big|(\theta,\psi), \*w_{t}\big) = c\left(\xi_t;\frac{2\|\*w_t^H\*a(\theta,\psi)\|^2}{\sigma^2}\right).
\end{equation}
\end{lemma}

We note that either observations $Z_t(P)$ or $Z_t(D)$ together with knowledge of the beamforming vector $\*w_t$ can be used to estimate the AoA $\Phi_t$. Furthermore, the quality of such an estimate can be directly controlled by the selection of the beamforming vector $\*w_t$. this forms the basis of our approach to posterior tracking and active selection of beamforming vectors. The specific communication protocol, active beamforming and dynamic posterior are discussed in Sect.~\ref{sect:comm_mobjour}.

\color{black}
\subsection{Mobility Model}
We consider a UAV mobility model where the AoA trajectory changes according to an independent increment process consisting of predictable and unpredictable (random) elements. That is, the AoA $\Phi_t$ evolves as

\begin{equation}
\label{phi_movement_mobjour}
\begin{aligned}
\Phi_{t+1} = \Phi_t + V + \*r_t + b \*q_t
\end{aligned} 
\end{equation}
where the known vector $V$ models predictable elements of mobility, for example an AoA position changing with a constant speed. The zero mean random vector $\*r_t\in \mathbb{R}^2$ models the unpredictable, yet zero-mean components such as drift or variations around the predictable mobility vector $V$. Lastly, the random vector $b \*q_t$ captures unpredictable components, such as a sudden jumps whose average size $b$ is known.

\subsection{Beamforming with a Codebook}
\begin{figure}
    \centering
    \includegraphics[width = 0.8\textwidth]{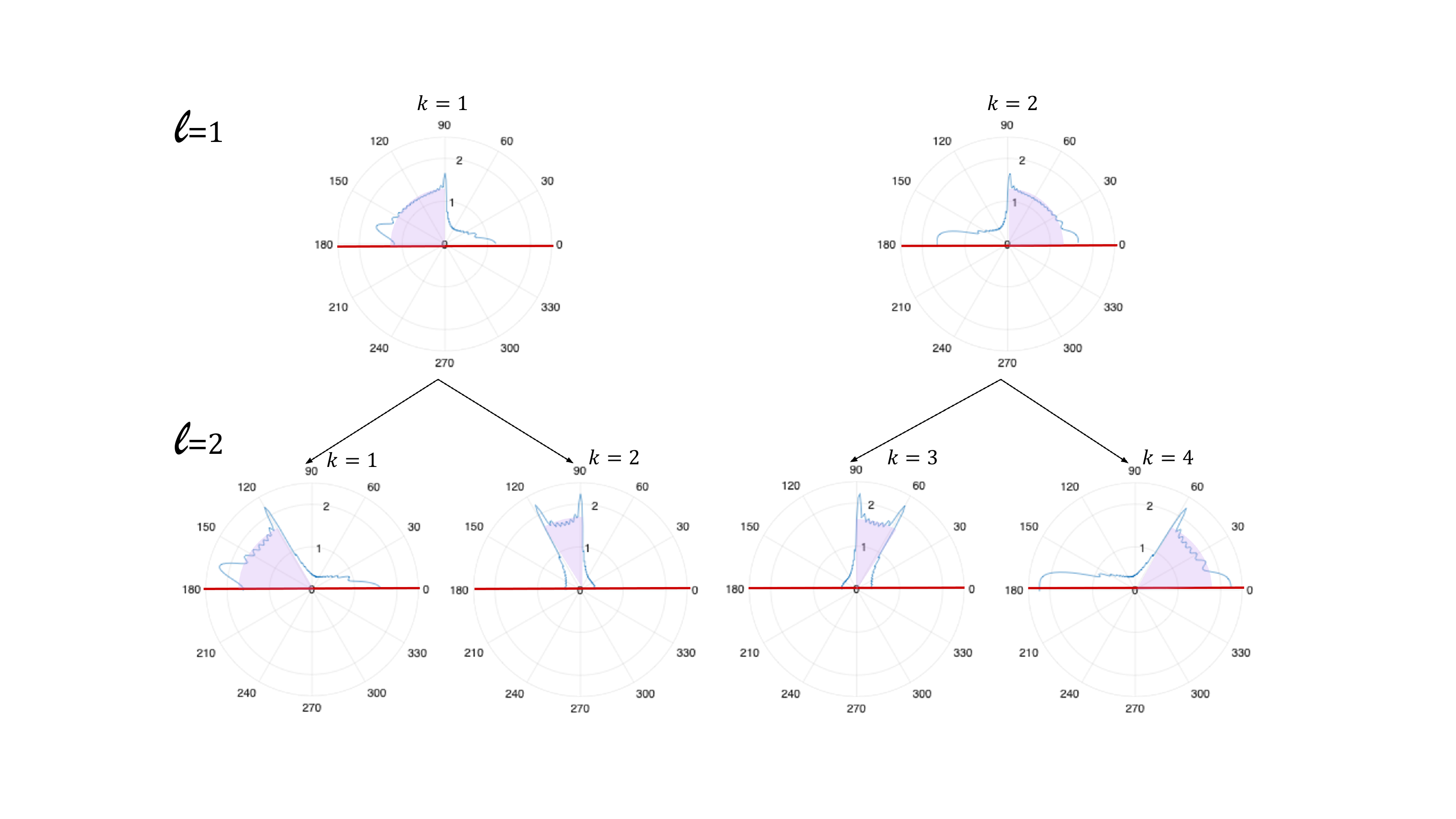}
    \caption{Illustration of the first 2 levels $l=1, \ l=2$ of the hierarchical beamforming codebook for the angular range of interest $[0, 180^{\circ}]$.}
    \label{fig:beamsfig_mobjour}
\end{figure}

At any given slot $t$, we are interested in selecting beam vectors for the receive beamforming. In the pilot phase, probing various beams allows the RX to learn about and ultimately track the AoA. In data transmission phase, the main goal is to select a beam vector $\*w_t$ whose main beam angular range includes the TX AoA $\Phi_t$ in order to receive and reliably detect a data sequence.
%We consider a stationary beamforming design policy as a causal (possibly random) mapping from past observations $\*z_{1:t} = [z_1(e_1), \ldots, Z_t(e_t)]$  and past beamforming $\*w_{1:t}$ to the beamforming vector: $\*w_{t+1} = \gamma(\*z_{1:t},\*w_{1:t})$.
In codebook-free methods, such as the deep learning approaches of \cite{sohrabi2021deep,khalili2021singleuser} for the static AoA case, the beamforming vector can be designed as $\*w_t\in \mathbb{C}^{MN}$ by formulating a constrained optimization problem. However, to reduce complexity it is common to limit the selection of the beamforming vector $\*w_t$ to a pre-designed beamforming codebook $\mathcal{W}^S$ with finite cardinality. In this work, we consider a multi-level codebook, $\mathcal{W}^S$, that has $S$ levels with $K_l$ vectors in each level $l\in S$ that partition the angular search space into contiguous sectors with increasing resolution $K_l<K_{l+1}$ and finest resolution $K_S = \Delta_a\times \Delta_e$ in azimuth and elevation. One such structure is achieved by Hierarchical codebooks which 
%we have previously considered in chapters~\ref{ch?}, and
are investigated for UAV communications in \cite{Yang2019} for small angle variations. Under a hierarchical codebook, for each level $l$, $K_l = 2^l$ beams partition the angular space evenly and each beam in the level $k \in \{1, \ldots,K_l\}$ has a main beam covering the range of angels $\mathcal{D}_l^{k}$. Furthermore, beamforming vectors are designed with the objective of near constant gain for intended directions, i.e. we have the following assumption of ideal beams.\footnote{We note that the assumption of ideal beams Assumption~\ref{asm:idealbeam_mobjour} simplifies our analysis, however, imperfect beamforming vectors obtained using a pseudo-inverse computation are fully accounted for in our simulations.} 
\begin{assumption} \label{asm:idealbeam_mobjour}
The beamforming vector $\*w\in \mathcal{W}^S$ at level $l$ covers a range of angles $\mathcal{D}_{l}^{k_t}$, and has constant beamforming power gain for any signal of AoA $\Phi_t\in D_{l}^{k_t}$ and rejects any signal outside of $D_{l}^{k_t}$, $i.e.$
\begin{equation} 
    \*w_t^H \*a(\Phi) = \begin{cases}  G_l, & \text{if } \Phi_t \in D_{l}^{k_t}\\
    0, & \text{if } \Phi_t \notin D_{l}^{k_t}
\end{cases}.
\end{equation}
\end{assumption}
We note that the assumption of ideal beams Assumption~\ref{asm:idealbeam_mobjour} simplifies our analysis, however, imperfect beamforming vectors obtained using a pseudo-inverse computation, illustrated in Fig.~\ref{fig:beamsfig_mobjour}, are fully accounted for in our simulations.
For ease of exposition, in this paper we also define the notation $\tilde{\*w}_t \in \{0,1\}^{\Delta_a\times \Delta_e}$ to be a binary matrix representation of the angular space of $\*w_t$. More specifically, the locations of 1's in $\tilde{\*w}_t$ indicate angular regions $\mathcal{D}_l^{k_t}$ covered by the beam $\*w_t$ in a corresponding to an $\Delta_a\times \Delta_e$ grid over the total angular space $[\theta_{\text{min}}, \theta_{\text{max}}]\times [\psi_{\text{min}}, \psi_{\text{max}}]$. 

\section{Proposed Algorithm}

We propose Alg.~\ref{algorithm_beamtracking_mobjour}, an adaptive communication scheme which allocates pilot training time adaptively and implements a procedure for actively and sequentially selecting beamforming vectors $\*w_{t+1}$ based on the accumulated belief about the AoA in the posterior vector $\boldsymbol{\pi}(t|t-1)$. Specifically, we implement $hiePM$ of \cite{ChiuJSAC2019} to actively select each $\*w_t$ via posterior matching, and build on this algorithm to incorporate an update and prediction step for evolving the posterior.
An overview of the proposed adaptive beamforming algorithm for mobile UAV is detailed in  Alg.~\ref{algorithm_beamtracking_mobjour}.
%Observations $z_{t}(e_t)$ can be thought of as the probing of certain angular locations, indicated by $\tilde{\*w}_t$, which give information about the presence or absence of the AoA in the angular space spanned by $\*w_t$. This enables us to actively learn $\Phi_t$ and sequentially select beamforming vectors whose angular width $\|\tilde{\*w}_t\|_0$ matches the accumulate belief about $\Phi_t$.
\subsection{Posterior Probability on AoA $\Phi_t$}
 Under Assumption \ref{assum:singlepath_mobjour}, the statistics of observations, both $Z_t(P)$ and $Z_t(D)$, depend on the estimate of channel state information $\hat{\*h}_t$, which is determined by a current estimate of the AoA $\hat{\Phi}_t$. In particular, each estimate provides beamforming $\*w_{t}$ which results in normalized beamforming gain: 
%\vspace{-3mm}
\begin{equation}\label{GBF_mobjour}
    G_{BF} = \mathbb{E}\big[\|\*w_{t}^{H} \*a(\Phi_t)\|^2\big].
\end{equation}
In other words, the quality of the established communication link as well as the utility of pilots over a period of time $t=[1:T]$ strongly depend on a method to robustly and continuously detect and track the AoA $\Phi_t$ for  $t=[1:T]$.
% The quality of both $Z_t(P)$ and $Z_t(D)$ depends on how well $\*w_t$ and the array vector $\*a(\Phi_t)$ are aligned, which is determined by the choice of $\*w_t$ according to the current knowledge of $\Phi_t$. 
Our proposed algorithm tracks the posterior belief on the AoA $\Phi_t$ in order to make decisions about the beamforming vector $\*w_t$ and pilot allocation $e_t$. More specifically, assuming a deterministic beam-selection strategy, we track the posterior probability over $\Phi_t$ given past observations $\*Z_{1:t} = [Z_1(e_1), Z_2(e_2), \ldots, Z_t(e_t)]$. %{\color{red}{\st{be  $\Phi_t\sim f_{\Phi_t|\*Z_{1:t},\*w_t}\left((\theta,\psi)|\.\xi_{1:t},\*w_t\right). $}}}
For computational feasibility, and under a slight abuse of notation, we define a discrete posterior $\.\pi(t|t)\in [0,1]^{\Delta_a\times \Delta_e}$,  where the $i=1,..., \Delta_a,  j=1,..., \Delta_e$ element is defined as 
\begin{equation}
\label{eq:prior_phi_mobjour}
\begin{aligned}
      \pi_{i,j}(t|t) := %\int_{\theta_i-\frac{\delta_a}{2}}^{\theta_i+\frac{\delta_a}{2}} \int 
    \mathbb{P}\left(\Phi_t \in  \Big[\theta_i-\frac{\delta_a}{2},\theta_i+\frac{\delta_a}{2}\Big)\times \Big[\psi_j-\frac{\delta_e}{2},\psi_j+\frac{\delta_e}{2}\Big)\ \bigg|\  \*z_{1:t}\right),
\end{aligned}
\end{equation}
denotes the conditional probability that the AoA $\Phi_t$ being in the angular range $ \Big[\theta_i-\frac{\delta_a}{2},\theta_i+\frac{\delta_a}{2}\Big)\times \Big[\psi_j-\frac{\delta_e}{2},\psi_j+\frac{\delta_e}{2}\Big)$, where $\theta_i = \theta_{\text{min}} + (i-\frac{1}{2})\delta_a$, and $\psi_j = \psi_{\text{min}} + (i-\frac{1}{2})\delta_e$, i.e. corresponding to bins with angular resolution $\delta_a=\frac{(\theta_{\text{max}}-\theta_{\text{min}})}{\Delta_a}$, and $\delta_e = \frac{(\psi_{\text{max}}-\psi_{\text{min}})}{\Delta_e}$ in azimuth and elevation\footnote{Note that the resolution of the discrete posterior ($\Delta_a \times \Delta_e$) matches the resolution of the finest-level beam vectors $k_S$.}. Furthermore, the probability of $\Phi_t$ being in the angular range covered by a beamforming vector $\*w_t$ (whose binary matrix angular representation $\tilde{\*w}_t$) is the sum of the posterior entries corresponding to the non-zero entries of $\tilde{\*w}_t$:
\begin{equation} \label{Bayes_update_pseudo_mobjour}
    \pi_{\tilde{\*w_t}}(t|t) :=  \sum\limits_{i,j}\tilde{\*w}_{t}(i,j)\pi_{i,j}(t|t).
\end{equation}

% We resolve the continuous AoA with a resolution ($\frac{1}{\delta_a} \times \frac{1}{\delta_e}$), i.e. a point estimate $\hat{\Phi}_t = (\hat{\phi}_{a,t},\hat{\phi}_{e,t})$ belongs to the discrete set where $\hat{\phi}_{a,t} = \{\theta_1, \theta_2,\ldots \theta_{1\slash \delta_a}\}$ and  $\theta_i = \theta_{\text{min}} + (i-\frac{1}{2})\times \delta_a \times (\psi_{\text{max}}-\psi_{\text{min}})$, and where $\hat{\phi}_{e,t} = \{\psi_1, \psi_2,\ldots \psi_{1\slash \delta_e}\}$ and  $\psi_i = \psi_{\text{min}} + (i-\frac{1}{2})\times \delta_e \times (\psi_{\text{max}}-\psi_{\text{min}})$. Let the probability over $\Phi_t$ given past observations $\*z_{1:t} = [Z_1(e_1), Z_2(e_2), \ldots, Z_t(e_t)]$
% be defined as $\.\pi(t|t)\in [0,1]^{\frac{1}{\delta_a} \times \frac{1}{\delta_e}}$, where
% \begin{equation}
% \label{eq:prior_phi_mobjour}
% \begin{aligned}
%       \Phi_t\sim \pi_{i,j}(t|t) := %\int_{\theta_i-\frac{\delta_a}{2}}^{\theta_i+\frac{\delta_a}{2}} \int 
%     \mathbb{P}{(\Phi_t = (\theta_i,\psi_j)|\*z_{1:t})}, \  i=1,..., 1\slash{\delta_a},  j=1,..., 1\slash{\delta_e}
% \end{aligned}
% \end{equation}
% and the probability of $\Phi_t$ being in the angular range covered by a beamforming vector $\*w_t$ (with angular span $\tilde{\*w}_t$) is the sum of the posterior entries corresponding to the non-zero entries of $\tilde{\*w}_t$:
% \begin{equation} \label{Bayes_update_pseudo_mobjour}
%     \pi_{\tilde{\*w_t}}(t|t) :=  \sum\limits_{i,j}\tilde{\*w}_{t}(i,j)\pi_{i,j}(t|t).
% \end{equation} 
% Our proposed algorithm utilizes this posterior belief on $\Phi_t$. 

\subsection{Integrated adaptive pilot selection}\label{sect:comm_mobjour}
As we saw in Sect.~\ref{signal_mobjour}, under Assumption~\ref{assum:singlepath_mobjour} the statistics of the signal $Z_t(e_t)$ depend on the choice of $\*w_t$ and channel state information $\Phi_t$, or more precisely on how $\*w_t$ aligns with $\Phi_t$. This means that the utility of both pilot and data transmission phases are influenced by how well $\Phi_t$ can be estimated. The information reward of the pilot signal is allowing the receiver to estimate the channel state information, in general, and in our setting specifically to provide noisy information about the AoA $\Phi_t$. For a given beamforming vector $\*w_t$, this information reward can be approximated by the mutual information between $Z_t(P)$ and $\Phi_t$, denoted as $I\big(\Phi_t; Z_t(P) \big| \*w_t,\.\pi(t|t-1)\big)$. 

\begin{figure}
    \centering
    \includegraphics[width = 0.6\textwidth]{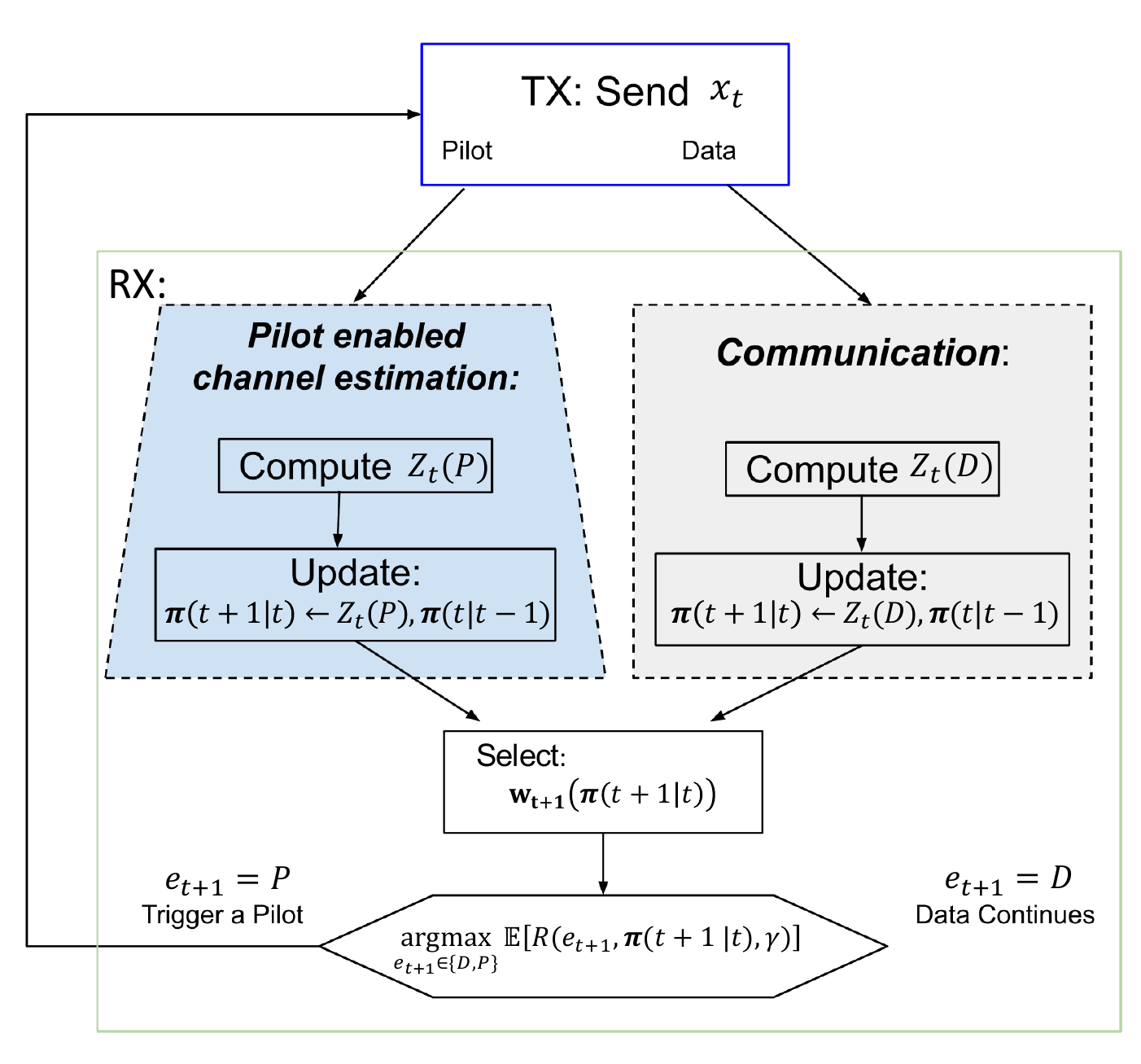}
    \caption{Overview of the proposed communication scheme. Adaptive pilot allocation is based on analysis of the mutual information and achievable spectral efficiency. }
    %by setting a threshold on the maximum posterior element $max(\theta_i,\psi_j)\boldsymbol{\pi}(t|t−1)\geq \rho^{\ast}$ required for transmission. When the posterior }
    \label{fig:overview_mobjour}
\end{figure}

On the other hand, the main benefit of the data transmission phase is the rate at which communication is possible. One way to measure this is via the expected spectral efficiency under a beamforming vector $\*w_t$, denoted as $S_t\big(D\big|\*w_t,\.\pi(t|t-1)\big)$. In addition to decoding $y_t$ and providing a non-zero communication rate (exploitation), the received signal $Z_t(D)$ can also be used to infer the AoA (exploration). In other words, the communication phase provides less efficient, yet non-zero mutual information about $\Phi_t$,  $I\big(Z_t(D);\Phi_t\big|\*w_t, \boldsymbol{\pi}(t|t-1)\big)$. Thus, the information reward of the pilot phase $I\big(Z_t(P);\Phi_t\big|\*w_t, \boldsymbol{\pi}(t|t-1)\big)$ can be traded off with the information exploitation reward of the data transmission phase. Specifically, let action $e_t\in \{D, P\}$ of triggering data (D) or pilot (P) transmission be determined by the following weighted analysis:
\begin{equation}\label{pilot_decision_mobjour}
\begin{aligned}
e_t &=\argmax_{e_t\in \{D, P\}} \mathbb{E}\big[R(e_t,\boldsymbol{\pi}(t|t-1),  \gamma)\big]\\
&= \argmax_{e_t\in \{D, P\}}I\big(\Phi_t; Z_t(e_t)\big|\*w_t, \boldsymbol{\pi}(t|t-1)\big) + \gamma S_t\big(e_t\big|\*w_t, \.\pi(t|t-1)\big)\\
\end{aligned}
\end{equation}
where $\gamma$ is an algorithm parameter trading off the potential to learn about the AoA with the maximum achievable spectral efficiency. Fig.~\ref{fig:overview_mobjour} is an overview of our pilot allocation approach.
 The mutual information terms can be formulated under Assumption~\ref{asm:idealbeam_mobjour} of ideal beams given in Lemma~\ref{MI_pilot_mobjour} below, while the communication reward is given in Lemma~\ref{SE_mobjour}.
\begin{lemma}
\label{MI_pilot_mobjour}
For the pilot phase $e_t=P$ the mutual information between the computed signal $Z_t(P)$ and the AoA is
\begin{equation}
\begin{aligned}
    &I\big(\Phi_t; Z_t(P)\big|\*w_t, \.\pi(t|t-1)\big) \\&= -\int_{-\infty}^{\infty}\int_{-\infty}^{\infty} f_{Z_t(P)|\*w_t}\big(\xi_t\big|\*w_t\big) \log{f_{Z_t(P)|\*w_t}\big(\xi_t\big|\*w_t\big)}d_{\mathfrak{R}(\xi_t)} d_{\mathfrak{I}(\xi_t)} -\log(\pi e \sigma^2),
\end{aligned}    
\end{equation}
where 
\begin{equation}
\begin{aligned}
   f_{Z_t(P)|\*w_t}\big(\xi_t\big|\*w_t\big)&=\boldsymbol{\pi}_{\tilde{\*w}_t}(t|t-1)  g\big(\xi_t;G_l\big)  +\big(1-\boldsymbol{\pi}_{\tilde{\*w}_t}(t|t-1) \big) g\big(\xi_t;0\big).\\
%   &=\boldsymbol{\pi}_{\tilde{\*w}_t}(t|t-1) \frac{1}{\pi \sigma^2} e^{-\frac{|\xi_t-G_l|^2}{\sigma^2}} +(1-\boldsymbol{\pi}_{\tilde{\*w}_t}(t|t-1) ) \frac{1}{\pi \sigma^2} e^{-\frac{|\xi_t|^2}{\sigma^2}}.\\
   \end{aligned}
\end{equation}
In the data transmission phase $e_t=D$, however, the mutual information between the computed signal $Z_t(D)$ and the AoA is
\begin{equation}
    I\big(\Phi_t; Z_t(D)\big|\*w_t\big)  = -\int_{-\infty}^{\infty} f_{Z_t(D)|\*w_t}\big(\xi_t\big|\*w_t\big) \log{f_{Z_t(D)|\*w_t}\big(\xi_t\big|\*w_t\big)}d{\xi_t} -h\big(Z_t(D)\big|\*w_t,\Phi_t\big), 
\end{equation}
where
\begin{equation}
\begin{aligned}
   f_{Z_t(D)|\*w_t}\big(\xi_t\big|\*w_t\big)&=\boldsymbol{\pi}_{\tilde{\*w}_t}(t|t-1)  c\left(\xi_t;\frac{2\|G_l\|^2}{\sigma^2}\right) +\big(1-\boldsymbol{\pi}_{\tilde{\*w}_t}(t|t-1) \big)  c\big(\xi_t;0\big)\\
   %&=\boldsymbol{\pi}_{\tilde{\*w}_t}(t|t-1)  \frac{1}{\sigma^2}e^{-(\frac{\xi_t+\|G_l\|^2}{\sigma^2})}\sum\limits_{k=0}^{\infty}\frac{(\frac{\xi_t\|G_l\|^2}{\sigma^4})^k}{(k!)^2}  +(1-\boldsymbol{\pi}_{\tilde{\*w}_t}(t|t-1) )  \frac{1}{\sigma^2}e^{\frac{-z_t}{\sigma^2}} , \ x\geq 0.\\
\end{aligned}
\end{equation}
and
\begin{equation}
\begin{aligned}
h\big(Z_t(D)\big|\*w_t,\Phi_t\big)=&-\boldsymbol{\pi}_{\tilde{\*w}_t}(t|t-1)\int_{-\infty}^{\infty}  c\left(\xi_t;\frac{2\|G_l\|^2}{\sigma^2}\right)\log{ c\left(\xi_t;\frac{2\|G_l\|^2}{\sigma^2}\right)}d{\xi_t}\\
& -\big(1-\boldsymbol{\pi}_{\tilde{\*w}_t}(t|t-1)\big)\int_{-\infty}^{\infty}   c\big(\xi_t;0\big) \log{c\big(\xi_t;0\big)}dz_t .\\ 
%=&-\boldsymbol{\pi}_{\tilde{\*w}_t}(t|t-1)\int_{-\infty}^{\infty}   \frac{1}{\sigma^2}e^{-(\frac{\xi_t+\|G_l\|^2}{\sigma^2})}\sum\limits_{k=0}^{\infty}\frac{(\frac{\xi_t\|G_l\|^2}{\sigma^4})^k}{(k!)^2} \log{\frac{1}{\sigma^2}e^{-(\frac{\xi_t+\|G_l\|^2}{\sigma^2})}\sum\limits_{k=0}^{\infty}\frac{(\frac{\xi_t\|G_l\|^2}{\sigma^4})^k}{(k!)^2} }d_{z_t}\\
%& -(1-\boldsymbol{\pi}_{\tilde{\*w}_t}(t|t-1))\int_{-\infty}^{\infty} \frac{1}{\sigma^2}e^{\frac{-z_t}{\sigma^2}}\log{ \frac{1}{\sigma^2}e^{\frac{-z_t}{\sigma^2}}}d\xi_t .\\
\end{aligned}
\end{equation}
where $\|G_l\|^2$ is the expected beamforming gain for a beam $\*w_t\in \mathcal{W}$ in level $l$ under Assumption~\ref{asm:idealbeam_mobjour}.
\end{lemma}
\begin{lemma}
\label{SE_mobjour}
For an action $e_t$, the maximum achievable spectral efficiency under a beamforming vector $\*w_t$ covering a range of angles $\mathcal{D}_{l}^{k_t}$, as indicated in the binary vector representation $\tilde{\*w}_t$, is given by
\begin{equation}\label{eq:data_rate_instant_mobjour}
     S_t\big(e_t\big|\*w_t,\.\pi(t|t-1)\big) =\boldsymbol{\pi}_{\tilde{\*w}_t}(t|t-1)  \log \left( 1+  \frac{\|G_l\|^2  }{\sigma^2}  \right) \mathds{1}_{e_t=D}.
\end{equation}
\end{lemma}
The proof of Lemmas~\ref{MI_pilot_mobjour} and~\ref{SE_mobjour} are given in the Appendix~\ref{Appendix_mobjour}, both following from Assumption~\ref{asm:idealbeam_mobjour}. 

\begin{remark}
The optimal choice of (\ref{pilot_decision_mobjour}) depends on the mutual information terms and thus requires an additional computational cost per iteration of the proposed algorithm if computed in an online manner. This cost can be reduced by assuming perfect beams and by approximately calculating the mutual information terms offline.
Under Assumption~\ref{asm:idealbeam_mobjour}, the theoretical beamforming gains can be used to approximate the mutual information terms of Lemma~\ref{MI_pilot_mobjour} offline for each level of the codebook $l \in S$ and for a range of input probabilities $\boldsymbol{\pi}_{\tilde{\*w}_t}(t|t-1) \in [0,1]$ ($n$ values chosen uniformly). As a result, an action (\ref{pilot_decision_mobjour}) can be chosen by comparing the pre-calculated mutual information terms for a given level $l$ and for a given input probability $\boldsymbol{\pi}_{\tilde{\*w}_t}(t|t-1)$ (by interpolating) and thus save on the online computational complexity. The additional computational cost per iteration is $O(2\log(\log(n))+2)$ if the mutual information terms are computed offline. 
\end{remark}
\color{black}
\subsection{Dynamic tracking and updating of the Posterior} 

To evolve the posterior, upon receiving a new observation $Z_t(e_t)$, $\.\pi(t|t-1)$ is updated according to Bayes Rule \cite{cover_book} and followed by a prediction step incorporating the AoA dynamics: 
\begin{equation}\label{Bayes_flow_mobjour}
    \.{\pi}(t+1|t) \leftarrow \.{\pi}(t|t) \leftarrow z_{t}(e_t), \.{\pi}(t|t-1).
\end{equation} 
To initialize the procedure of (\ref{Bayes_flow_mobjour}) we assume a uniform posterior $\boldsymbol{\pi}(1|0)$ at $t=1$, i.e. no prior knowledge about the AoA is required. The first step in (\ref{Bayes_flow_mobjour}) is a Bayesian posterior update calculated as:
\begin{equation}\label{eq:JPM_discrete_mobjour}
\pi_{i,j}(t|t) =\frac{f_{Z_{t}(e_t)|\Phi_t,\*w_t}\big(\xi_t\big|(\theta_i,\psi_j), \*w_{t}\big) \pi_{i,j}(t|t-1)}
    {\sum\limits_{i'=1}^{\Delta_a}\sum\limits_{j'=1}^{\Delta_e}   f_{Z_{t}(e_t)|\Phi_t, \*w_t}\big(\xi_{t}\big|(\theta_{i'},\psi_{j'}), \*w_{t}\big) \pi_{i',j'}(t|t-1)}.
\end{equation}
where $f_{Z_{t}(e_t)|\Phi_t, \*w_t}\big(\xi_t\big|(\theta_i,\psi_j), \*w_{t}\big)$ is the conditional probability density function of ${Z_{t}(e_t)}$ when beamforming vector $\*w_t$ is selected and the AoA $\Phi_t \in  \Big[\theta_i-\frac{\delta_a}{2},\theta_i+\frac{\delta_a}{2}\Big)\times \Big[\psi_j-\frac{\delta_e}{2},\psi_j+\frac{\delta_e}{2}\Big)$. %\footnote{Recall $\Phi_t = (\theta_i, \psi_j)$ denotes the the AoA $\Phi_t$ being in the angular range $ [\theta_i-\frac{\delta_a}{2},\theta_i+\frac{\delta_a}{2})\times[\psi_j-\frac{\delta_e}{2},\psi_i+\frac{\delta_e}{2})$.}. 
In the pilot phase, the Bayes posterior update~(\ref{eq:JPM_discrete_mobjour}) is computed utilizing (\ref{condprob_pilot_mobjour}) for all $t$ where $e_t = P$. In the data transmission phase of communication (\ref{condprob_pilot_mobjour}) does not apply since the RX does not have knowledge of the transmitted data. However, the received power measurement $Z_t(D)$ and corresponding conditional probability, defined in Lemma~\ref{cond_prob_data_mobjour}, can be used as a proxy for computing the Bayes Rule. Applying this power only Bayesian update will enable some learning of the AoA without a pilot. 
That is, the calculated power of a data signal $Z_t(D)$ can be used to an extent for confirming (increasing probability of angles covered by $\*{\tilde{w}}_t$) or rejecting (decreasing probability of angles covered by $\*{\tilde{w}}_t$) the current use of the tracking beam $\*w_t$. As a result, the duration of the data transmission phase may be extended so long as a posterior continues to increase for the selected beam. %Our proposed strategy for determining the tracking duration and thereby the pilot allocation is discussed in Sect.~\ref{sect:comm_mobjour}.
%However, in the case of a serious misalignment, the beam will be rejected and drastic posterior refinement will likely only be achieved using the pilot phase. Occasionally, after any number of transmission slots, the receiver can determine unfit tracking reliability and again trigger the pilot phase. Our proposed pilot allocation strategy is discussed in Sect.~\ref{sect:comm_mobjour}.

\subsection{Active and sequential selection of beamforming vectors}
 We propose to implement an active beamforming policy $\gamma$ based on Bayesian posterior updates and predictions, and which achieves sequential refinement of uncertainty on the dynamic AoA $\Phi_t$. Specifically, we implement $hiePM$ of \cite{ChiuJSAC2019} to actively select each $\*w_t$ via posterior matching. A beamforming vector at either level $l$ or $l+1$ is selected for the next beamforming slot based on the accumulated belief around $\Phi_{t}$ as described by the prediction posterior probability  $\boldsymbol{\pi}(t|t-1)$, which is a sufficient statistic. In other words, $\*w_{t} \big(\boldsymbol{\pi}(t|t-1)\big)$ is chosen as the $k_{t}^{\ast th}$ beam in level $l_{t}^{\ast}$ covering the angular range $\mathcal{D}_{l_{t}^{\ast}}^{k_{t}^{\ast}}$ where:
\begin{equation}\label{hiePM_wt_mobjour}
\left[l_{t}^{\ast},k_{t}^{\ast}\right] = \argmin_{[l,k]} \[|\pi_{\tilde{\*w}_{[l,k]}}(t|t-1) - \frac{1}{2}\]|
\end{equation}
and $\tilde{\*w}_{[l,k]}$ denotes by the binary vector representation of the $k^{th}$ beam in level $l$.

\begin{remark} 
We note that our proposed approach has computational requirements from the pilot allocation, the selection policy (\ref{hiePM_wt_mobjour}), the Bayesian posterior update (\ref{eq:JPM_discrete_mobjour}), and the one-step prediction update. 
While the first steps of the proposed algorithm can be generalized as above, in order to incorporate mobility information, the one-step posterior update from  $\boldsymbol{\pi}(t|t)$ to $\boldsymbol{\pi}(t+1|t)$ is formulated 
according to a particular movement model (\ref{phi_movement_mobjour}). Even so, the computation complexity of our proposed algorithm is dominated by the cost of the posterior update (\ref{eq:JPM_discrete_mobjour}) at each beamforming slot $t$. The worst case cost is $O\left(\Delta_a\times \Delta_e\right)$ if each element is updated individually. However, this can be reduced to $O\left(log(\Delta_a\times \Delta_e)\right)$ by considering the geometric constraints  on the hierarchical contiguous codebook elements \cite{Sungen2020}.  
%Furthermore, computation of  (\ref{Bayes_flow_mobjour}) will depend on knowledge of the one-step prediction function of the belief vector from  $\boldsymbol{\pi}(t|t)$ to $\boldsymbol{\pi}(t+1|t)$, which is formulated for a particular movement model (\ref{phi_movement_mobjour}) since it incorporates available mobility information. 
% In section~\ref{allexamples_mobjour}, we formulate the one-step prediction function for a set of general examples to give a sense of how statistics of the stochastic mobility are incorporated into such function. 
\end{remark}

\begin{algorithm}[] \caption{Active beam tracking for mobile AoA }
 \label{algorithm_beamtracking_mobjour}
 \textbf{Input}: target resolution $({\Delta_a,\Delta_e})$, tracking quality parameter $\gamma$, codebook $\mathcal{W}^S$, T (total duration), Markov mobility model (\ref{phi_movement_mobjour})\\
 \textbf{Output}: Beam vector $\*w_{t}\in \mathcal{W}$ and pilot allocation $e_t\in \{P,D\}$%, and estimate of the AoA $\hat{\Phi}_t$ up to a resolution $\delta$ for each slot $t$
 \\
 \textbf{Initialization}: Set $\boldsymbol{\pi}(1|0)$ to be uniform, i.e. $\pi_{i,j}(t|t-1) = \frac{1}{\Delta_a\Delta_e}$\\
 \While{$t \leq T$}
 {\# Beam Selection~(\ref{hiePM_wt_mobjour}): hierarchical posterior matching with variable width beams as a function of $\.\pi(t|t-1)$:
    $\tilde{\*w}_{[l_{t}^{\ast},k_{t}^{\ast}]}$ where:
    \begin{equation*}\label{hiePM_wt_mobjour}
    [l_{t}^{\ast},k_{t}^{\ast}] = \argmin_{[l,k]} \[|\pi_{\tilde{\*w}_{[l,k]}}(t|t-1) - \frac{1}{2}\]|. 
    \end{equation*}\\
   
  \# Allocate Pilot or Data: select based on the trade-off (\ref{pilot_decision_mobjour}) between the information and communication rewards: 
  
    \begin{gather*}
    e_t = \argmax_{e_t\in \{D, P\}}I\big(\Phi_t; Z_t(e_t)\big|\*w_t, \.\pi(t|t-1)\big) + \gamma S_t\big(e_t\big|\*w_t,\.\pi(t|t-1)\big)
    \end{gather*}
   \# Receive and compute observation: \\
   \indent \# Obtain received signal at the output of the RF chain: $y_t =\sqrt{P_T} \*w_t^H \*h x_t + \*w_t^H\*n_{t}$\\
   
   \eIf{$e_t = P$}{
    \# Compute: $Z_t = Z_t(P)$ according to (\ref{eq:obsv_mobjour})\\
    }
    {\# Compute: $Z_t = Z_t(D)$ according to (\ref{eq:obsv_data_mobjour})\\
     }
      \# Posterior update by Bayes' Rule (\ref{eq:JPM_discrete_mobjour})
     \begin{gather*} 
     \label{eq:Bay_mobjour}
         \.\pi(t|t) \leftarrow Z_{t}, \.\pi(t|t-1)
     \end{gather*} \\
     \# Posterior one-step prediction based on mobility model  (\ref{phi_movement_mobjour})
     \begin{gather*} 
     \label{eq:onesteppred_mobjour}
         \.\pi(t+1|t) \leftarrow   \.\pi(t|t)
     \end{gather*} \\
     %\# Estimate of AoA: $\hat{\Phi}_{t+1} = \argmax_{(\theta_i,\psi_j)} \pi_{i,j}(t+1|t)$\\
     }
\end{algorithm}

\section{Numerical Results}
In this section we illustrate the proposed tracking scheme under some examples of stochastic mobility and analyze the performance with extensive simulations. 

\subsection{Stochastic Mobility Models}\label{allexamples_mobjour}
We consider the special case where we reduce the AoA point estimates to the 2-D angular domain $\hat{\Phi}_t=\hat{\phi}_t\in[\theta_{\text{min}},\theta_{\text{max}}]$, with target resolution $\Delta$ and corresponding $\delta = \frac{(\theta_{\text{max}}-\theta_{\text{min}})}{\Delta}$, and consider three examples of stochastic mobility in the form of  (\ref{phi_movement_mobjour}). 

% We consider three movement models that fall into two broad categories: predictable and unpredictable mobility. Predictable mobility is considered to be a known angular movement increment, like a constant velocity, where $\phi_{t+1}$ can be calculated from $\phi_t$. We consider unpredictable mobility to be modeled by stochastic processes and we look at two examples: Gaussian angular movements and Bernoulli angular jumps. We show that predictable mobility is equivalent to no movement when the predictable mobility can be accounted for in the beam selection. We show that stochastic mobility is much harder to handle, but can be efficiently tracked by incorporating side information into the prediction step of (\ref{Bayes_flow_mobjour}). We provide the one-step prediction formulation (\ref{Bayes_flow_mobjour}) for these three example cases of UAV movement.
To provide intuition we consider three special cases: 
\subsubsection{Predictable Angular Movement}
Consider the AoA to change only according to a fixed angular velocity:

\begin{equation} \label{phi_movement_constantv_mobjour}
\begin{aligned}
\phi_{t+1} = \phi_t + V
\end{aligned} 
\end{equation}
where $V = \nu \delta$ summarizes the constant angular movement of $\nu \delta$ per time slot. In this work we assume that $V$ is known, however, a small preamble to determine an unknown angular velocity is easily implemented as in \cite{Huang2020}. Intuitively, for integer values of $\nu$ the corresponding one-step prediction in (\ref{Bayes_flow_mobjour}) is:

\begin{equation}
    \pi_{i}(t+1|t)= \pi_{i-\nu}(t|t),
\end{equation}
and for $|\nu|<1$

\begin{equation}
    \pi_{i}(t+1|t)= (1-\nu)\pi_{i}(t|t)+\nu\pi_{i+sign(\nu)}(t|t).
\end{equation}

\begin{remark}\label{Remarkvel_mobjour}
 The one-step posterior prediction $\.{\pi}(t+1|t)$ is a shifted version of the posterior update $\.{\pi}(t|t)$, this is easiest to see for integers $\nu$, where this is a simple horizontal translation. For any predictable mobility the one-step prediction will result in a shifting or deterministic rearranging of the posterior $\.{\pi}(t|t)$. As a result, we can apply the same fundamental limits as \cite{ChiuJSAC2019} in terms of estimation error probability and time required to obtain a robust initial estimate of $\hat{\phi}_t$.
\end{remark}

\subsubsection{Gaussian Angular Movement}
Consider the mobility scenario where the AoA changes with Gaussian angular movements due to small intractable position changes on the UAV such as with small drones.
That is, the AoA evolves as:

\begin{equation}\label{Moving_phi_noise_mobjour}
\begin{aligned}
\phi_{t+1} = \phi_t + \*r_t
\end{aligned} 
\end{equation}
where $\*r_t$ is an i.i.d. zero mean Gaussian with variance $\sigma_{\phi}^2$. Intuitively, this is a good model for small uncertainties about direction or vibrations. Note that in case of a misalignment event, the cumulative movement results in a linear growth in uncertainty. The corresponding one-step prediction in (\ref{Bayes_flow_mobjour}) is: 
\begin{equation}
    \pi_i(t+1|t)=\left\langle\.\pi(t|t), \*g_{[\theta_i,\sigma_{\phi}^2]}\right\rangle,
\end{equation}
where $\*g_{[\theta_i,\sigma_{\phi}^2]}\in[0,1]^{\Delta}$ is a probability mass function obtained from a quantized and truncated Gaussian random variable $x\sim \mathcal{N}(\theta_i,\sigma_{\phi}^2)$ with resolution $\delta$. Equivalently, each element is given as:

\begin{equation}
g_{[\theta_i,\sigma_{\phi}^2]}(n) \propto \mathbb{P}\left[\theta_n-\frac{\delta}{2}\leq x<\theta_n+\frac{\delta}{2}\right],
\end{equation}
for $n=\{1, \ldots, \Delta\}$, and $\sum \*g_{[\theta_i,\sigma_{\phi}^2]} = 1$.

\subsubsection{Bernoulli Angular Jumps}
Next, consider the case where the AoA can incur a large random jump from one beamforming slot to another. We assume the  AoA moves according to

\begin{equation}\label{phimovement_jumps_mobjour}
\begin{aligned}
\phi_{t+1} = \phi_t + b \*q_t 
\end{aligned} 
\end{equation}
where $b= \beta \delta$ is a known probable jump size and $\*q_t$ is a Bernoulli random variable with parameter $p$, where $p$ is the probability of a jump. Note that in this case the mean change in the AoA, given as $bp$, is the predictable mobility component while $r_t=b \*q_t - bp$ is the mean zero unpredictable and random component. This mobility pattern can occur for example in the cases of blockage or sudden changes in velocity. This mobility model is often difficult to handle because the random movement almost certainly will cause an outage if the beamforming is not updated quickly. As a result of such jumps, existing tracking methods will likely trigger a reset due to invalid tracking (i.e. not meeting a minimum tracking quality). If the jump is small enough, a Kalman filtering strategy may try to update the estimate on the state $\phi_t$ based on observations $y_t$ and catch up. 
In contrast, our approach is to conservatively account for the uncertainty about $\phi_t$ preemptively, by widening the posterior in the prediction step, that is by accounting for the likelihood of jumps by increasing the posterior probability in probable jump locations. 
For integer estimates of the jump size $\beta$, the one-step prediction in (\ref{Bayes_flow_mobjour}) can be specified as: 

\begin{equation}
    \pi_{i}(t+1|t)= (1-p) \ \pi_{i}(t|t) + p\pi_{i-\beta}(t|t).
\end{equation}
In general, the mobility model for a target AoA may differ from the ones considered here. The aim of this work is to introduce the idea of incorporating mobility information into the the selection of beamforming vectors for tracking, especially for the cases where the movement may be stochastic, in order to robustly handle outage scenarios. The main idea is to use prior information and adapt to the uncertainty by widening and shrinking the beam width $\|\tilde{\*w}_t\|_0$ in response to a misalignment rather than forcing a reset protocol.
% \begin {table}[H]
% \caption{One-Step Posterior Prediction for Markov Movements }\label{table_onesteps_mobjour}
% \vspace{-5mm}
% \begin{center}
% {\renewcommand{\arraystretch}{1.5}
% \begin{tabular}{ |p{2cm}|p{3cm}|p{5cm}| } 
% \hline
%  &Movement $\phi_{t+1} = $& \hfil One-step Predict $\pi_i(t+1|t)=$\\
% \hline
% \hline
% \hfil Static &\hfil $\phi_t$&\hfil  $\pi_{i}(t|t)$ \\ 
% \hline
% \hfil Constant &\hfil $\phi_t +\nu \delta \pi$& \hfil $\pi_{i-\nu}(t|t),\  \nu$ integer\\ 
% & &  $(1-\nu)\pi_{i}(t|t)+\nu\pi_{i\pm1}(t|t),\  |\nu|<1$ \\ 
% \hline
% \hfil Gaussian & \hfil $\phi_t + \omega,$ & \hfil $\langle\.\pi(t|t), \*g_{[\theta_i,\sigma_{\phi}^2]}\rangle$ \\ 
% &\hfil $ \omega\sim\mathcal{N}(0,\sigma_{\phi}^2)$ &\\
% \hline
% \hfil Jumps &\hfil  $\phi_t + b\*q,  $ & \hfil $ (1-p)\pi_{i}(t|t) + p\pi_{i-\beta}(t|t)$ \\ 
% &\hfil $\*q\sim \text{Bern}(1-p) $&\\
% \hline
% \end{tabular}}
% \end{center}
% \end {table}

\subsection{Simulation Scenario} \label{sims_mobjour}
Next, we analyze the performance of the proposed communication scheme under the mobility examples discussed above.
We consider a simulation scenario where the RX is equipped with N=32 antennas in a uniform linear array. We utilize the hierarchical beamforming codebook of \cite{Alkhateeb2014} where vectors in level $l\in S = log_2(\Delta)$, for $\Delta=64$,\footnote{Correspondingly, for $\phi_t \in [-180^{\circ}, 0]$ we have narrowest angular resolution parameter $\delta = \frac{(\theta_{\text{max}}-\theta_{\text{min}})}{\Delta} = \frac{180^
{\circ}}{64}$} have angular width $\|\tilde{\*w}_t\|_0 = \frac{\Delta}{2^l}$ that is half the size of the prior level.
This is approximately achieved via a pseudo inverse approximation. The resulting beams are slightly imperfect with reduced gain in angles further from the center beam directions, however, these effects are fully accounted for in our numerical simulations. 
% and restricts point estimates of the AoA, such that $1 \slash \delta=64,$ i.e. $\hat{\phi}_t \in \{\theta_1,\ldots, \theta_{64}\}$ where the codebook $\mathcal{W}^{\log_2{\frac{1}{\delta}}}$ has levels $l \in \{1,\ldots, 6\}$. We assume the starting point of the AoA $\phi_0$, lies uniformly in the angular space, i.e. $\boldsymbol{\pi}(1|0) = \boldsymbol{\delta}_{\frac{1}{\delta} \times 1}$, and randomize $\phi_0\in[-\pi, 0]$ to fully account for cases on the edges of the angular space. 
The transmitted data symbols have a minimum energy $\|x_t\|\geq1$, which we obtain by using a QPSK constellation \cite{Tse2005}.
We focus on relative comparisons to existing tracking strategies in terms of the performance measures of normalized beamforming gain and pilot overhead for a given SNR, although physical properties corresponding to the considered SNR values (like distance, cell size, and bandwidth) can be defined as in (Fig.6 in \cite{ChiuJSAC2019}). More specifically, we compare to the following prior works:
\begin{itemize}
    \item The Extended Kalman Filtering algorithm of \cite{Va2016_GSIP} selects beams based on tracking estimates of the state $\hat{\phi}_t$ with extended Kalman filter updates. The pilot allocation is determined by a threshold on the mean squared error, i.e. %sqrt(MSE)
    $\sqrt{\mathbb{E}\left[| \phi_t-\hat{\phi}_t|^2\right]}\geq\frac{BW}{2}$ half beam width. When tracking is no longer valid, a reset is triggered and the exhaustive beams are used again in order to obtain another estimate. 
    \item In the dynamic pilot insertion algorithm of \cite{Huang2020}, beams are selected based on predictions of AoA made using an estimated velocity. The pilot allocation is determined by a threshold on the normalized receive power $\Big(\frac{\mathbb{E}[\|\*w_t^H\*a(\phi_t)\|^2]}{\mathbb{E}[\|\*w_{\tau}^H\*a(\phi_{\tau})\|^2]}\geq P_{min}$\Big) where $\tau$ is the first transmission slot and $t>\tau$.
    \item The beam tracking strategy of \cite{Yang2019} employs beams from a certain level of the hierarchical codebook $\*w_t \in \mathcal{W}^S$ (we consider either narrow ($l=6$) or wide ($l=5$) beams). After an initial estimate of the AoA is obtained, subsequent recurring pilot phases consists of scanning only the neighboring local beams from the current estimate. %Wide beams can be leveraged for handling small angle variations, with the caveat of decreased signal strength (or reduced maximum transmission distance). 
    The training frequency, i.e. tracking duration $\tau_{max}$ between pilot phases, is determined according to the channel coherence time.
    %adaptively by estimating the channel coherence time and by considering the distance to TX (our metric of operating SNR captures the parameter of distance). 
    %For the following simulations we assume perfect estimates of the channel coherence time - which remains fixed, and thus consider this recurring local algorithm under either narrow ($l=6$) or wide ($l=5$) beams.  
\end{itemize}

All of the strategies we compare to utilize the pilot phase for acquiring an aligned estimate $\hat{\phi}_{T_E}$ (or re-estimation if tracking thresholds are not met) before switching to the transmission phase. In the absence of a better solution, and under the constraint of a single RF chain, we apply an exhaustive search over all candidate beams in order to obtain an estimate $\hat{\phi}$ for these algorithms. The duration of the exhaustive search $T_E$ will depend on the beam width of the candidate beams; $T_E=\Delta$ for narrow beams and $T_E=2^l$ for beams from any other level $l$ of the codebook $\mathcal{W}^S$. The tracking thresholds $\sqrt{MSE}$ and $P_{min}$, and the tracking duration $\tau_{max}$ of the strategy \cite{Yang2019} may be optimized for a given SNR or coherence time. In the following simulations we assume perfect exhaustive search estimates $\hat{\phi}_{T_E} = \phi_{T_E}$ and optimize the tracking parameters empirically as best as we can in order to compare performance.

\subsection{Impact of the parameter $\gamma$}
\begin{figure}[]
    \centering
    \includegraphics[width = 1\textwidth]{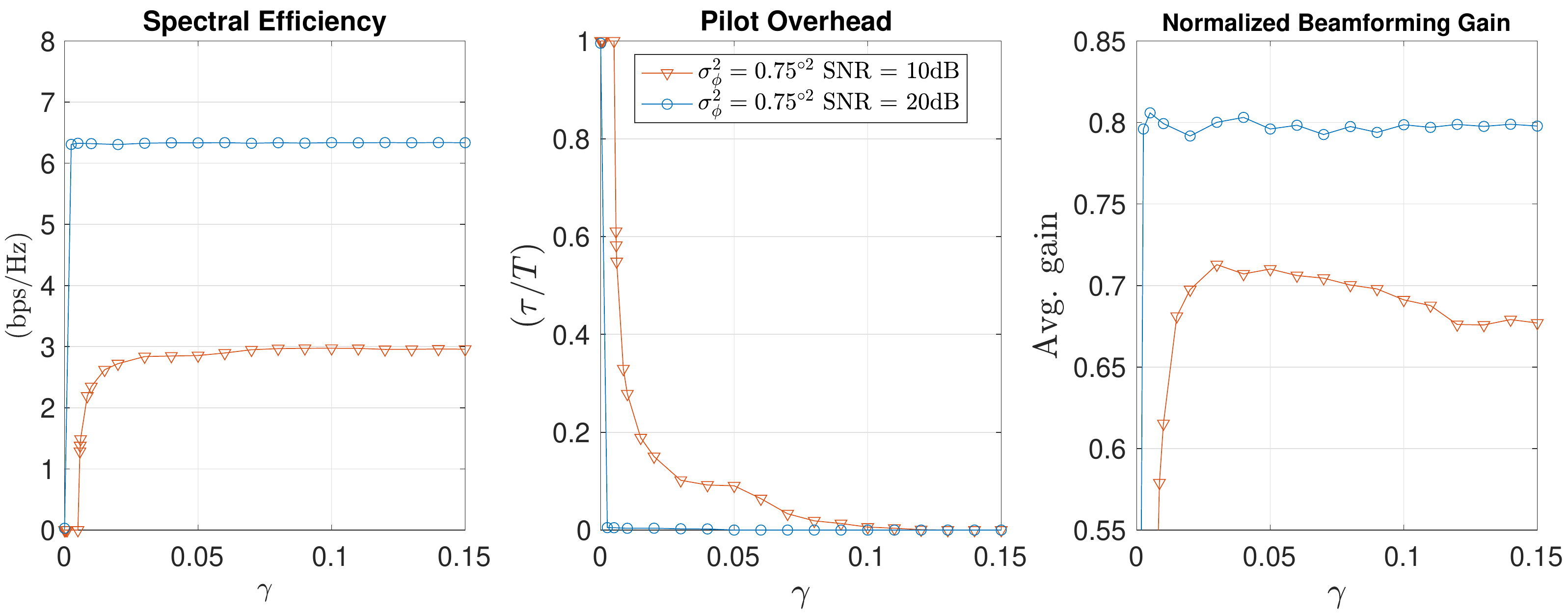}
    \caption{We investigate performance as a function of the choice of $\gamma$ under the mobility model of Gaussian angular movements with variance ($\sigma^2_{\phi} = 0.75^{\degree2}$).}
    \label{analyze_gamma_mobjour}
\end{figure}
First, we discuss the impact of the parameter $\gamma$ on the performance measures of pilot overhead $\sum_{t=1}^T \mathds{1}_{e_t=P}$, average received beamforming gain 
$\expect*{G_{BF}}$ of (\ref{GBF_mobjour}), and spectral efficiency \\ $\mathbb{E}\Big[S_t\big(e_t\big|\*w_t, \.\pi(t|t-1)\big)\Big]$. The total time frame is set arbitrarily large, $T>>\Delta$ at $T = 500$ beamforming slots.
% \begin{equation}
%     \mathbb{E}\Bigg [ \frac{T-\tau}{T}\log{1+\frac{P|\*w_t^H\*a(\phi_t)|^2}{\sigma^2} } \Bigg ]
% \end{equation}
% where $\tau$ is the total number of pilot slots in all pilot phases over a time frame $T$. 
In Fig.~\ref{analyze_gamma_mobjour} we plot the performance under the mobility model of Gaussian angular movements with variance $\left(\sigma^2_{\phi} = 0.75^{\degree2}\right)$ as a function of the parameter $\gamma$ for SNR = $\frac{P}{\sigma^2} =10$~dB, and SNR = $\frac{P}{\sigma^2} = 20$~dB. We see that selecting a large $\gamma$ %larger than a threshold (dictated by the SNR, e.g $\gamma>0.01$ for SNR$= 20$~dB) 
improves the spectral efficiency and reduces pilot overhead until these values saturate. This indicates that given high enough signal power, active and dynamic learning of the AoA in the data  transmission phase is sufficient for maintaining alignment and results in high spectral efficiency even as the pilot overhead is reduced to $0$ (in this example $\gamma>0.1$). Alternatively, to maximize the average beamforming gain $\gamma$ may be optimized for each SNR. In this example, for $20$~dB SNR $\gamma^{\ast} = 0.005$, and for $10$~dB SNR $\gamma^{\ast} = 0.03$.

\subsection{Comparative Analysis of the Results}

\subsubsection{Predictable Angular Movement}

\begin{figure}[]
    \centering
    \includegraphics[width = 1\textwidth]{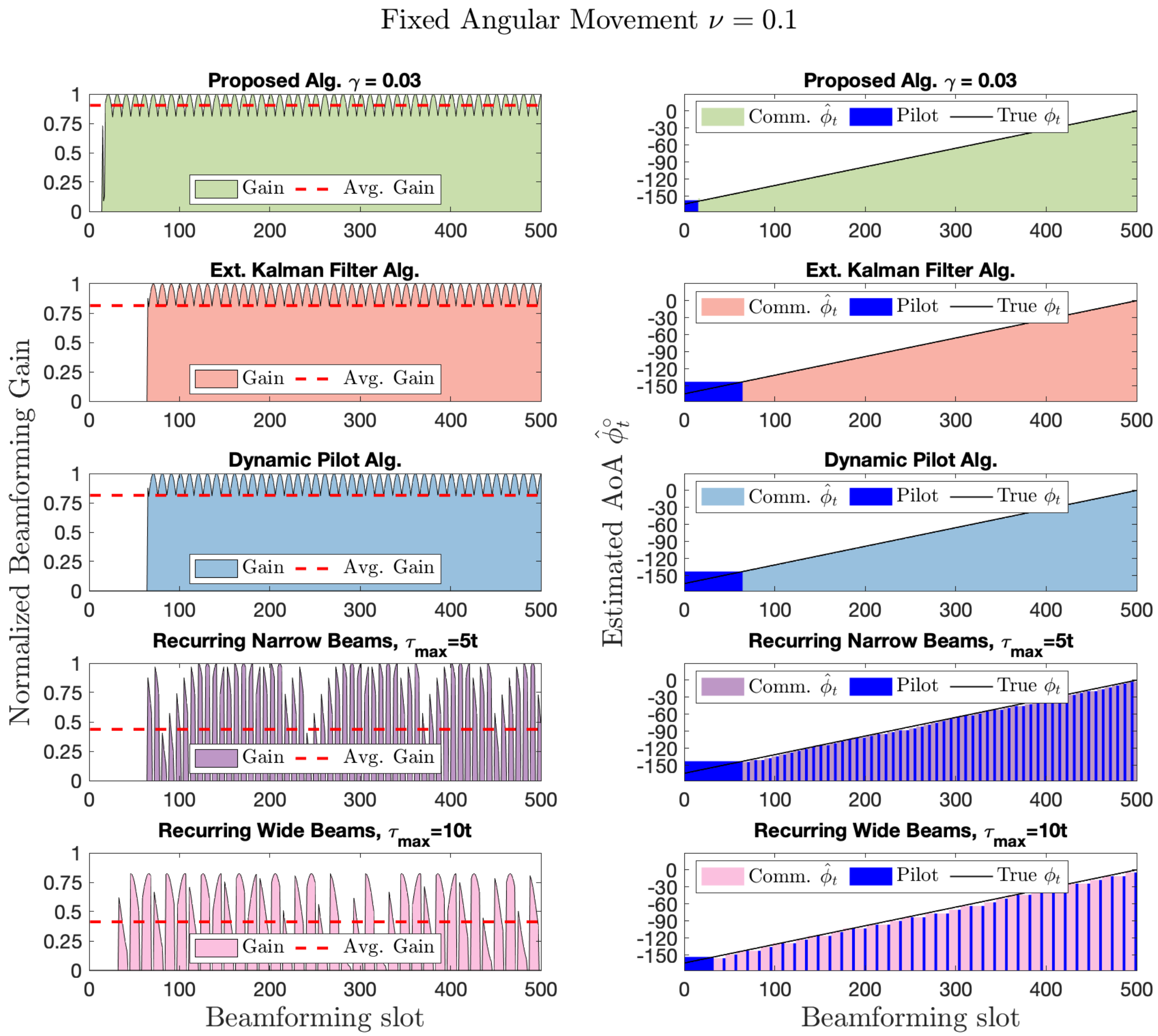}%velM01_gamma_v2_copy.pdf
    \caption{Normalized beamforming gain (\ref{GBF_mobjour}) at 10dB SNR for constant angular movement (\ref{phi_movement_constantv_mobjour}) with $\nu = 0.1$ at $10$dB SNR. For the proposed algorithm $\gamma = 0.03$. On the right, the estimated AoA is compared to the true AoA and pilot allocation is shown. 
    %obtained by the proposed algorithm compared to the extended Kalman filtering algorithm of \cite{Va2016_GSIP}, and the dynamic pilot insertion algorithm of \cite{Huang2020}.
    }
    \label{Moving_phi_continuousvel_mobjour}
\end{figure}

Next, we get a sense of performance of the proposed communication algorithm compared to existing approaches in terms of tracking quality by analyzing the achieved normalized beamforming gain (\ref{GBF_mobjour}) over time. %A beamforming gain equal to 0 indicates a pilot phase slot, thus also illustrating the pilot overhead incurred by each strategy. 
% We assume the conditions of perfect exhaustive search and empirically optimize tracking parameters for \cite{Va2016_GSIP}, \cite{Huang2020}, and \cite{Yang2019}.
In Fig.~\ref{Moving_phi_continuousvel_mobjour} we show an AoA trajectory example under the mobility model with predictable angular movement (\ref{phi_movement_constantv_mobjour}) with $\nu = 0.1$, i.e. increments  $V = 0.1 \delta$, at $10$dB SNR. The normalized beamforming gains achieved are shown on the left, and corresponding AoA estimates and pilot allocation are shown on the right. For the proposed algorithm, the AoA estimate $\hat{\phi}_t$ is defined as the main lobe pointing direction of the selected beam $\*w_t$.
% The proposed strategy and the strategies of \cite{Va2016_GSIP} and \cite{Huang2020} allocate pilots adaptively, albeit in different manners, and are all able to remain in the communication phase so long as the mobility model does not change, i.e. fully predictable mobility. Additionally, under this fixed mobility model we note these algorithms are able to accurately adjust their AoA estimates based on a known mobility and can successfully account for this predictable movement. The Neighborhood search strategy of \cite{Yang2019} updates estimates of $\hat{\phi}_t$ based on periodical monitoring of the neighboring beams, thus incurs some pilot overhead even in this predictable mobility scenario. 
In this example, all strategies are able to remain in the data transmission phase so long as the mobility is known or estimated correctly, i.e. fully predictable mobility.
% With accounting for this predictable mobility \cite{Yang2019} will perform similarly to the other algorithms considered (although we do not show those results in this figure). 
We notice that using wider beams only, rather than recurring narrow beams, in the Neighborhood search strategy of \cite{Yang2019} results in slightly reduced maximum beamforming gain but achieves longer tracking duration, as expected.
% Overall, none of the algorithms are significantly affected by predictable mobility with a known mobility other than 
Slight dips in the beamforming gain are caused when the AoA lies near the edge of a selected beam or by using wider beams. In essence, Fig.~\ref{Moving_phi_continuousvel_mobjour} shows that under predictable mobility only, there is little difference in the performance by the various algorithms considered.
Fig.~\ref{Moving_phi_continuousvel_mobjour} also highlights a strength of the proposed algorithm in obtaining an initial estimate of the AoA quickly and reliably, thereby initiating the transmission phase with tracking significantly more quickly than the compared to applying the exhaustive search for this initial alignment. As a result of this and subsequent short-duration pilot phases, the proposed algorithm also achieves the highest average beamforming gain.

\subsubsection{Gaussian Angular Movement} 
\begin{figure}[]
    \centering
    \includegraphics[width = 1\textwidth]{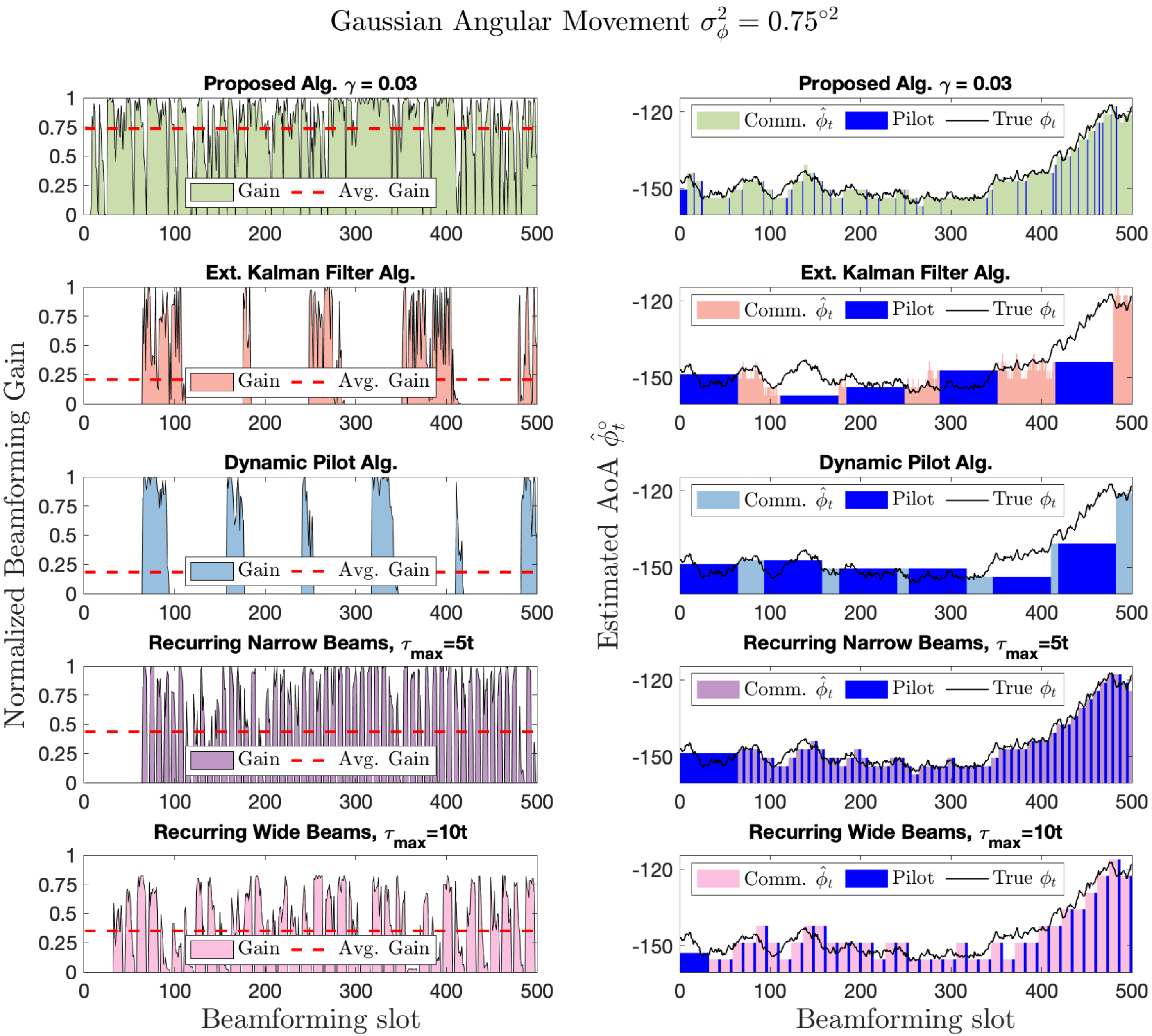}%SIG_gamma_copy.pdf
    \caption{Normalized beamforming gain (\ref{GBF_mobjour}) at 10dB SNR for Gaussian Movement (\ref{Moving_phi_noise_mobjour}) with $\sigma_{\phi}^2 =0.75^{\degree2}$. For the proposed algorithm $\gamma=0.03$. On the right, the estimated AoA is compared to the true AoA and pilot allocation is shown. 
    }   
    \label{Moving_phi_randnoise_mobjour}
\end{figure}
Next, we analyze the more interesting cases of mobility with stochastic elements, starting with the scenario of Gaussian angular movements (\ref{Moving_phi_noise_mobjour}). 
In Fig.~\ref{Moving_phi_randnoise_mobjour} we look a very high mobility case of incremental Gaussian movements with variance $\sigma_{\phi}^2 = 0.75^{\degree 2}$. We plot an example AoA trajectory along with the corresponding estimates and beamforming gains achieved over time by the algorithms considered. 
The algorithms of \cite{Va2016_GSIP} and \cite{Huang2020} have strict quality thresholds that trigger re-estimation when the random movement overcomes the predictable movement and the performance drops enough that the tracking is deemed invalid. 
Under the high mobility scenario considered in this example, these pilot phases are triggered frequently and a full scan over the potential beamforming vector pairs is costly resulting in a large amount of time spent in re-estimating  the AoA (pilot phase) compared to tracking (in the data transmission phase)\footnote{Here we note that both the algorithms of \cite{Va2016_GSIP} and \cite{Huang2020} suggest reducing the overhead of the re-estimation phases according to current CSI estimates after initial alignment, however, no clear strategies for doing this are provided.}. The strategy of frequently analyzing local neighboring beams \cite{Yang2019}, whether with narrow or wide beams, improves on the other strategies due to the shorter pilot phases (scanning only neighboring beams). 
Our proposed algorithm recovers the benefit of high gains achieved by the tracking strategies of \cite{Va2016_GSIP} and \cite{Huang2020}, as well as the benefit of lower overhead incurred by searching locally based on prior estimates \cite{Yang2019} providing an overall more efficient strategy. Combined, our sequential beam selection and adaptive pilot allocation yield sustained larger gains overtime. %, high gains are sustained even when the parameter $\gamma = 0.13$ is set so high that no pilots are used. 
Ultimately, our proposed algorithm enables efficient beam tracking for high mobility by incorporating mobility information into the sequential beam selection with variable width beams and into the pilot allocation strategy. 

\subsubsection{Bernoulli Angular Jumps} 
\begin{figure}[]
    \centering
    \includegraphics[width = 1\textwidth]{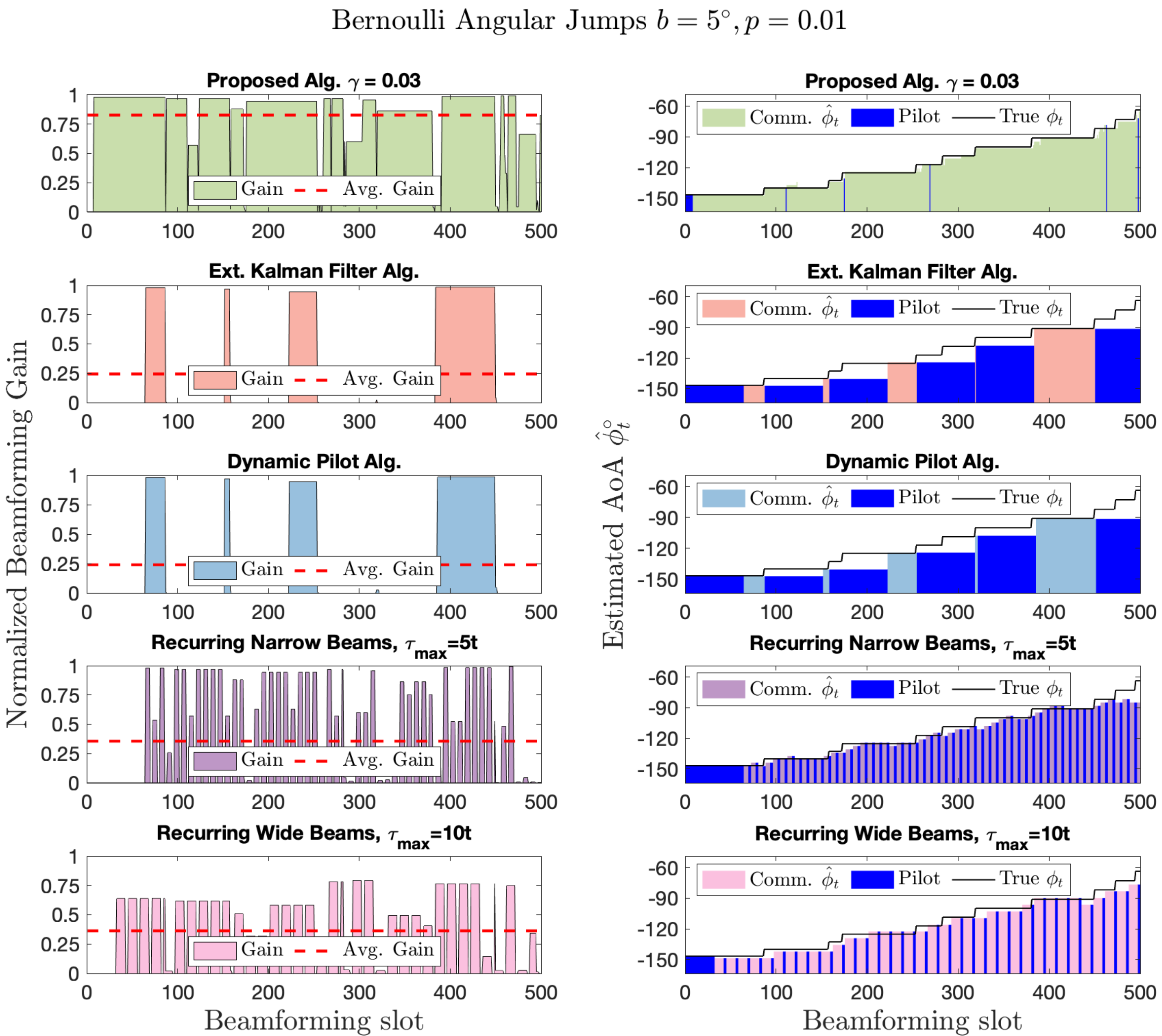}%Jump5_099_gamma_copy.pdf
    \caption{Normalized beamforming gain (\ref{GBF_mobjour}) at 10dB SNR for Bernoulli jumps (\ref{phimovement_jumps_mobjour}) with jump size $b= 5\degree$ and probability $p = 0.01$. For the proposed algorithm $\gamma=0.03$. On the right, the estimated AoA is compared to the true AoA and pilot allocation is shown. %obtained by the proposed algorithm compared to the extended Kalman filtering algorithm of \cite{Va2016_GSIP}, and the dynamic pilot insertion algorithm of \cite{Huang2020}.
    }   
    \label{Moving_phi_randjumps_mobjour}
\end{figure}
% \begin{comment}
% \begin{figure}[]
%     \centering
%     \includegraphics[width = 0.45\textwidth]{Jumps_V2.eps}
%     \caption{Random jumps (\ref{phimovement_jumps}) with jump size $b= 5 \degree$ and probability $p = \{0.01,0.05,0.20\}$: Normalized beamforming gain (\ref{GBF}) at 10dB SNR obtained by the proposed algorithm, the Extended Kalman filtering algorithm of \cite{Va2016_GSIP}, and the Dynamic pilot insertion algorithm of \cite{Huang2020}.}   
%     \label{Moving_phi_randjumps}
% \end{figure}
% \end{comment}
Lastly, we analyze the stochastic mobility model of occasional Bernoulli jumps (\ref{phimovement_jumps_mobjour}).
Such random jumps are very difficult to handle on two fronts. First, existing alignment schemes operate under the assumption quasi-static AoA at least for the duration of the initial alignment phase, and will struggle to obtain correct initial AoA estimates if a jumps occurs in this phase. Second, even if a robust AoA estimate is obtained successfully, the duration of data transmission phase (active tracking) will largely depend on the frequency of the jumps (the entropy of the mobility model), since each jump is likely to cause an outage that requires re-estimation of the AoA.\footnote{We note that the Extended Kalman Filter updates of the algorithm of \cite{Va2016_GSIP} are designed for state estimates of AoA under Gaussian noise, thus are not specifically designed for this mobility. Updates adapted for this mobility model are not immediately obvious to us, thus no such extensions are considered in this paper.}
In Fig.~\ref{Moving_phi_randjumps_mobjour} we plot an AoA trajectory and corresponding estimates and beamforming gain performance comparisons for each of the algorithms considered.
As expected, under this unpredictable mobility the adaptive algorithms of \cite{Va2016_GSIP} and \cite{Huang2020} respond to sudden jumps by triggering the pilot phase for re-estimation of the AoA. We note that the pilot phase frequency and corresponding duration of the communication phases is correlated to the frequency and spread of jumps, where consecutive jumps can severely shorten the data transmission phase. A similar effect occurs for the recurring local search algorithm of \cite{Yang2019}, where consecutive jumps cause the algorithm under narrow beams be misaligned more severely\footnote{Under unpredictable mobility with large jumps, these recurring or periodical algorithms would benefit from scanning a larger radius around the current CSI estimates, although though these extensions are not considered here.}. The local search under wide beams is more robust to frequent jumps, although estimates are less precise and the beamforming gain suffers. On the other hand, the proposed algorithm recovers quickly after a jump occurs due to the embedding of mobility information into dynamic evolution of the posterior and the active selection of beams with various widths.  %We see that the beamforming gain is only slightly reduced even when a large $\gamma = 0.13$ is selected and no pilots are used.  

\section{Conclusion}
 We consider the problem of active and dynamic sequential tracking of the CSI for robust communications at mmWave frequencies and above. We are interested in tracking stochastic movements, which may be especially critical in systems of communication between mobile UAV. Existing beam tracking communication schemes implement approaches which require lengthy or too frequent re-estimation pilot phases in response to outages, and require long coherence times to maintain tracking quality. We propose a communication scheme that consists of a strategy for actively selecting beam vectors, a method for evolving the posterior, and an  adaptive pilot allocation based on information and communication rewards. Our proposed beamforming algorithm incorporates mobility information into the sequential selection of beams based on a posterior belief vector that is updated upon receiving observations. 
To allocate pilot slots adaptively we continuously analyze the expected information and communication rewards of each (pilot and communication) phase via analysis of the mutual information and spectral efficiency terms. 
This adaptive allocation strategy is driven by a weighting parameter $\gamma$ which can be chosen based on the performance measures of pilot overhead, average received SNR, and spectral efficiency, but is not too terribly sensitive for the SNR of 10~dB or 20~dB. Although we provide a general formulation for our algorithm that can be adapted to any stochastic mobility model, this paper assumes knowledge of the model parameters when the mobility is unknown a learning algorithm that can provide the stochastic mobility model may be used to complement this work. In our numerical analysis, we study a selection of 2-D Markov mobility models and provide the closed form equations for posterior updates and predictions for these examples. Our Numerical results show improved performance over existing strategies in terms of sustained beamforming gain, enabling tracking for movements with larger entropy.

\section{Appendix}\label{Appendix_mobjour}
\subsection{Proof of Lemma~\ref{cond_prob_data_mobjour}}
We assume $\alpha_t=1$ and $P_T=1$, and recall that $Z_t(D) =\|y_t\|^2=$ $\mathfrak{R}(y_t)^2+\mathfrak{I}(y_t)^2$. Furthermore, $\mathfrak{R}(y_t)^2\sim \mathcal{N}\left(\mathfrak{R}\big(\*w_t^H \*a(\Phi_t)x_t\big),\frac{\sigma^2}{2}\right)$ and  $\mathfrak{I}(y_t)\sim \mathcal{N}\left(\mathfrak{I}\big(\*w_t^H \*a(\Phi_t)x_t\big), \frac{\sigma^2}{2}\right)$. 
Thus conditioned on $\Phi_t=(\theta,\psi), $  and beamforming vector $\*w_{t}$, $Z_t(D)$ is the sum of two Gaussian random variables squared, which by definition gives that  $Z_t(D)\sim\chi^2(k,\lambda_t)$ follows a non-central chi-squared probability distribution function scaled by the variance $\frac{\sigma^2}{2}$ with $k=2$ degrees of freedom and time-varying non-centrality parameter 
\begin{equation}
\begin{aligned}
\lambda_t&=\frac{\Big(\mathfrak{R}\big(\*w_t^H \*a(\theta,\psi)x_t\big)\Big)^2}{\sigma^2 \slash 2}+\frac{\Big(\mathfrak{I}\big(\*w_t^H \*a(\theta,\psi)x_t\big)\Big)^2}{\sigma^2 \slash 2}\\
&=\frac{2\|\*w_t^H \*a(\theta,\psi)x_t\|^2}{\sigma^2} \\
&\geq \frac{2\|\*w_t^H \*a(\theta,\psi)\|^2}{\sigma^2}
\end{aligned}
\end{equation}
We approximate the conditional probability of $Z_t(P)$ with the worst possible symbol energy  
$\|x_t\|^2=1$, i.e by setting $\lambda_t=\frac{2\|\*w_t^H \*a(\theta,\psi)\|^2}{\sigma^2}$. 

\subsection{Proof of Lemma~\ref{MI_pilot_mobjour}}
\subsubsection{For the pilot phase, where $e_t=P$}

Let $\alpha_t=1$, and $P_T=1$, thus recall the received measurement model $Z_t(P)$ for the pilot phase is modeled as: 
%can delete this later or reference chapter 
\begin{equation}
\begin{aligned}
    Z_t(P) &{=}  \*w_t^H \*a(\Phi_t) + \*w_t^H \*n_{t}.
\end{aligned}
\end{equation} 
Under Assumption~\ref{asm:idealbeam_mobjour}, for $e_t = P$ with beamforming vector $\*w_t$ in level $l_t=l$, the conditional probability density function of $Z_t(P)$ is
\begin{equation}
\begin{aligned}
   f_{Z_t(P)|\*w_t}\big(\xi_t\big|\*w_t\big)&= \int_{-\infty}^{\infty}f_{Z_t(P)|\*w_t,\Phi_t}\big(\xi_t\big|\*w_t,(\theta,\psi)\big)d{(\theta,\psi)}\\
   &= \prob{\Phi_t\in \mathcal{D}_{l}^{k_t}} f_{Z_t(P)|\*w_t,\Phi_t}\big(\xi_t\big|\*w_t,\Phi_t\in \mathcal{D}_{l}^{k_t}\big)\\
   &+\prob{\Phi_t\notin \mathcal{D}_{l}^{k_t}} f_{Z_t(P)|\*w_t,\Phi_t}\big(\xi_t\big|\*w_t,\Phi_t\notin \mathcal{D}_{l}^{k_t}\big) \\
   &=\boldsymbol{\pi}_{\tilde{\*w}_t}(t|t-1)  \mathcal{CN}(G_l,\sigma^2)   +\big(1-\boldsymbol{\pi}_{\tilde{\*w}_t}(t|t-1) \big)   \mathcal{CN}\left(0,{\sigma^2}\right)\\
   &=\boldsymbol{\pi}_{\tilde{\*w}_t}(t|t-1)     \frac{1}{\pi \sigma^2} e^{-\frac{\|\xi_t-G_l\|^2}{\sigma^2}} +\big(1-\boldsymbol{\pi}_{\tilde{\*w}_t}(t|t-1) \big) \frac{1}{\pi \sigma^2} e^{-\frac{\|\xi_t\|^2}{\sigma^2}}. \\
   &=\boldsymbol{\pi}_{\tilde{\*w}_t}(t|t-1)  g\big(\xi_t;G_l\big)  +(1-\boldsymbol{\pi}_{\tilde{\*w}_t}(t|t-1) ) g\big(\xi_t;0\big)\\
\end{aligned}
\end{equation}
Additionally, for normalized beams $\|\*w_t\|^2=1$ 
 $\eta_t = \*w_t^H \*n_{t} \sim \mathcal{CN}(0,\sigma^2)$, which yields
\begin{equation}
\begin{aligned}
    h(\eta_t) &= h\big(\mathfrak{R}(\eta_t)\big)+ h\big(\mathfrak{I}(\eta_t)\big)\\
    &=\frac{1}{2}\log(2\pi e \frac{\sigma^2}{2})+\frac{1}{2}\log(2\pi e \frac{\sigma^2}{2})\\
    &=\log(\pi e \sigma^2).
\end{aligned}
\end{equation}
The mutual information term for the pilot phase action $e_t = P$ of (\ref{pilot_decision_mobjour}) is 
\begin{equation}
\begin{aligned}
    &I\big(\Phi_t; Z_t(P)\big|\*w_t, \.\pi(t|t-1)\big) \\
    &= h\big(Z_t(P)\big|\*w_t\big)-h\big(Z_t(P)\big|\*w_t,\Phi_t\big)\\
    &=h\big(Z_t(P)\big|\*w_t\big) - h(\eta_t)\\
    &=h\big(Z_t(P)\big|\*w_t\big) -\log(\pi e \sigma^2)\\
    & = -\int_{-\infty}^{\infty}\int_{-\infty}^{\infty} f_{Z_t(P)|\*w_t}\big(\xi_t\big|\*w_t\big) \log{f_{Z_t(P)|\*w_t}\big(\xi_t\big|\*w_t\big)}d{\mathfrak{R}({\xi_t})} d{\mathfrak{I}({\xi_t})} -\log(\pi e \sigma^2).\\
\end{aligned}
\end{equation}

\subsubsection{For the data phase, where $e_t=D$}
By Lemma~\ref{cond_prob_data_mobjour} and under Assumption~\ref{asm:idealbeam_mobjour} $Z_t(D)\sim\chi^2(k =2,\lambda_t \geq \frac{2\|\*w^H_t\*a(\Phi_t)\|^2}{\sigma^2})$ scaled by the variance $\frac{\sigma^2}{2}$. Under the worst case assumption of $\|x_t\|^2=1$, i.e. by setting $\lambda_t= \frac{2\|\*w^H_t\*a(\Phi_t)\|^2}{\sigma^2}$,
the distribution of $z_t$ conditioned on a beam $\*w_t$ of level $l_t=1$ is approximated as
%(reference for this is \cite{Moser_2020})
\begin{equation}
\begin{aligned}
   &f_{Z_t(D)|\*w_t}(\xi_t|\*w_t)\\
   &= \int_{-\infty}^{\infty}f_{Z_t(D)|\*w_t, \Phi_t}(\xi_t|\*w_t,(\theta,\psi))d_{(\theta,\psi)}\\
   &= \prob{\Phi_t\in \mathcal{D}_{l}^{k_t}} f_{Z_t(D)|\*w_t, \Phi_t}(\xi_t|\*w_t,\Phi_t\in \mathcal{D}_{l}^{k_t})+\prob{\Phi_t\notin \mathcal{D}_{l}^{k_t}}f_{Z_t(D)|\*w_t, \Phi_t}(\xi_t|\*w_t,\Phi_t\notin \mathcal{D}_{l}^{k_t}) \\
   \end{aligned}
   \end{equation}
   \begin{equation*}
\begin{aligned}
   &=\boldsymbol{\pi}_{\tilde{\*w}_t}(t|t-1)   \chi^2(2,\lambda_t=\frac{2\|G_l\|^2}{\sigma^2})  +(1-\boldsymbol{\pi}_{\tilde{\*w}_t}(t|t-1) )   \chi^2(2,\lambda_t=0)\\
%   &=\boldsymbol{\pi}_{\tilde{\*w}_t}(t|t-1)    \frac{1}{2(\sigma^2/2)}e^{-(\frac{z_t}{2(\sigma^2/2)}+\frac{\lambda_t}{2})}\sum\limits_{i=0}^{\infty}\frac{(\frac{z_t \lambda_t}{4(\sigma^2/2)})^i}{(i!)^2} +(1-\boldsymbol{\pi}_{\tilde{\*w}_t}(t|t-1) ) \frac{1}{2(\sigma^2/2)}e^{\frac{-z_t}{2(\sigma^2/2)}} \\
   &=\boldsymbol{\pi}_{\tilde{\*w}_t}(t|t-1)  \frac{1}{\sigma^2}e^{-(\frac{\xi_t}{\sigma^2}+\frac{\lambda_t}{2})}\sum\limits_{k=0}^{\infty}\frac{(\frac{\xi_t \lambda_t}{2\sigma^2})^k}{(k!)^2}  +(1-\boldsymbol{\pi}_{\tilde{\*w}_t}(t|t-1) )  \frac{1}{\sigma^2}e^{\frac{-\xi_t}{\sigma^2}} \\
   &=\boldsymbol{\pi}_{\tilde{\*w}_t}(t|t-1)  \frac{1}{\sigma^2}e^{-(\frac{\xi_t-\|G_l\|^2}{\sigma^2})}\sum\limits_{k=0}^{\infty}\frac{(\frac{\xi_t\|G_l\|^2}{\sigma^4})^k}{(k!)^2}  +(1-\boldsymbol{\pi}_{\tilde{\*w}_t}(t|t-1) )  \frac{1}{\sigma^2}e^{\frac{-\xi_t}{\sigma^2}}. \\
\end{aligned}
\end{equation*} 
Additionally, we have the conditional entropy 
\begin{equation}
\begin{aligned}
&h\big(Z_t(D)\big|\*w_t,\Phi_t\big) \\
%&= \sum\limits_{\mathcal{X}}\prob{X_t = x|\*w_t}h(z_t|X_t = x,\*w_t)\\
&=\prob{\Phi_t\in \mathcal{D}_{l}^{k_t}}h\big(Z_t(D)\big|\*w_t,\Phi_t\in \mathcal{D}_{l}^{k_t}\big)+\prob{\Phi_t\notin \mathcal{D}_{l}^{k_t}}h\big(Z_t(D)\big|\*w_t,\Phi_t\notin \mathcal{D}_{l}^{k_t}\big)\\
%&=\boldsymbol{\pi}_{\tilde{\*w}_t}(t|t-1)  h(z_t|X_t = G_lx(e_t))+(1-\boldsymbol{\pi}_{\tilde{\*w}_t}(t|t-1)  )h(z_t|X_t = 0)\\
&{=}\boldsymbol{\pi}_{\tilde{\*w}_t}(t|t-1)  h\big(Z_t(D)\big|\Phi_t\in \mathcal{D}_{l}^{k_t}\big)+\big(1-\boldsymbol{\pi}_{\tilde{\*w}_t}(t|t-1)  \big)h(|\eta_t|^2)\\
&= -\boldsymbol{\pi}_{\tilde{\*w}_t}(t|t-1)\int_{-\infty}^{\infty} \chi^2\left(2,\frac{2\|G_l\|^2}{\sigma^2}\right)  \log{\chi^2\left(2,\frac{2\|G_l\|^2}{\sigma^2}\right) }d_{\xi_t} \\
&\ \ \ -\big(1-\boldsymbol{\pi}_{\tilde{\*w}_t}(t|t-1)\big)\int_{-\infty}^{\infty} \chi^2(2,0)  \log{\chi^2(2,0) }d{\xi_t} \\
&=-\boldsymbol{\pi}_{\tilde{\*w}_t}(t|t-1)\int_{-\infty}^{\infty}   \frac{1}{\sigma^2}e^{-\left(\frac{\xi_t-\|G_l\|^2}{\sigma^2}\right)}\sum\limits_{k=0}^{\infty}\frac{\left(\frac{\xi_t\|G_l\|^2}{\sigma^4}\right)^k}{(k!)^2} \log{  \frac{1}{\sigma^2}e^{-\left(\frac{\xi_t-\|G_l\|^2}{\sigma^2}\right)}\sum\limits_{k=0}^{\infty}\frac{\left(\frac{\xi_t\|G_l\|^2}{\sigma^4}\right)^k}{(k!)^2} }d_{\xi_t}\\
&\ \ \ -\big(1-\boldsymbol{\pi}_{\tilde{\*w}_t}(t|t-1)\big)\int_{-\infty}^{\infty}   \frac{1}{\sigma^2}e^{\frac{-\xi_t}{\sigma^2}} \log{  \frac{1}{\sigma^2}e^{\frac{-\xi_t}{\sigma^2}}}d_{\xi_t}.\\
\end{aligned}
\end{equation}
where $\eta_t = \*w_t^H\*N_{t}\sim{CN}(0,\sigma^2)$ from our model (\ref{eq:obsv_data_mobjour}).
Thus, the mutual information term for an action $e_t=D$ of  (\ref{pilot_decision_mobjour}) is
\begin{equation}
\begin{aligned}
    I\big(\Phi_t; Z_t(D)\big|\*w_t, \.\pi(t|t-1)\big)  &= h\big(Z_t(D)\big|\*w_t\big)-h\big(Z_t(D)\big|\*w_t,\Phi_t\big)\\
    &= -\int_{-\infty}^{\infty} f_{Z_t(D)|\*w_t}\big(\xi_t\big|\*w_t\big) \log{f_{Z_t(D)|\*w_t}\big(\xi_t\big|\*w_t\big)}d{\xi_t} -h\big(Z_t(D)\big|\*w_t,\Phi_t\big) \\
\end{aligned}
\end{equation}

\subsection{Proof of Lemma~\ref{SE_mobjour}}
For an action $e_t=D$, the average spectral efficiency under a beamforming vector $\*w_t$ covering a range of angles $\mathcal{D}_{l}^{k_t}$, as indicated in the binary vector representation $\tilde{\*w}_t$, is given as
\begin{equation}
\begin{aligned}
     S_t\big(D\big|\*w_t,\.\pi(t|t-1)\big)&= \expect*{ \log \[( 1+  \frac{\| \*w_t^H\*a(\Phi_t)x_t\|^2 }{\sigma^2}  \])}. \\
\end{aligned}
\end{equation}  
On the other hand $S_t\big(P\big|\*w_t,\.\pi(t|t-1)\big)=0$ for the pilot phase.
Thus, for any action $e_t$ the maximum achievable spectral efficiency under a beamforming vector $\*w_t$ covering a range of angles $\mathcal{D}_{l}^{k_t}$, as indicated in the binary vector representation $\tilde{\*w}_t$, can be written as
\begin{equation}
\begin{aligned}
     S_t\big(e_t\big|\*w_t,\.\pi(t|t-1)\big)&= \expect*{ \log \[( 1+  \frac{\| \*w_t^H\*a(\Phi_t)x_t\|^2 }{\sigma^2}  \])\mathds{1}_{e_t=D} } \\
\end{aligned}
\end{equation}  
\begin{equation*}
\begin{aligned}
     &\ \ \ = \prob{\Phi_t \in \mathcal{D}_{l}^{k_t}}  \log \[( 1+  \frac{\| G_l x_t \|^2 }{\sigma^2}  \])  \mathds{1}_{e_t=D}\\
\end{aligned}
\end{equation*}  
\begin{equation*}
\begin{aligned}
     &\ \ \ =\boldsymbol{\pi}_{\tilde{\*w}_t}(t|t-1)  \log \[( 1+  \frac{\|G_l\|^2  }{\sigma^2}  \]) \mathds{1}_{e_t=D}
    \\
\end{aligned}
\end{equation*}
where $\|G_l\|^2$ is the expected beamforming gain for a beam $\*w_t$ in level $l_t=l$ under Assumption~\ref{asm:idealbeam_mobjour}.

\bibliographystyle{IEEEtran}%IEEEtran IEEEbib
%\vspace{-4mm}
\bibliography{./refs}

\end{document}